\documentclass[prd,a4paper,aps,showpacs,twocolumn,floats,nofootinbib]{revtex4}
\usepackage{epsfig}
\usepackage{graphicx}
\usepackage{color}
\usepackage{amssymb}

\newcommand{\bn}{{\mathbf n}}

\newcommand{\de}{\delta}

\newcommand{\la}{\lambda}

\newcommand{\om}{\omega}
\newcommand{\si}{\sigma}
\newcommand{\Si}{\Sigma}

\newcommand{\ra}{\rightarrow}

\newcommand{\be}{\begin{equation}}
\newcommand{\ee}{\end{equation}}
\newcommand{\gsim}{\stackrel{>}{\sim}}

\newcommand{\bea}{\begin{eqnarray}}
\newcommand{\eea}{\end{eqnarray}}
\newcommand{\bean}{\begin{eqnarray*}}
\newcommand{\eean}{\end{eqnarray*}}

\newcommand{\HH}{{\mathcal H}}

\begin{document}

\title{Vector and Tensor Contributions to the Luminosity Distance}

\author{Enea Di Dio  and Ruth Durrer}
\affiliation{D\'epartement de Physique Th\'eorique and Center for Astroparticle Physics, 
Universit\'e de 
Gen\`eve\\ 24 quai Ernest 
Ansermet, CH--1211 Gen\`eve 4, Switzerland}


\date{\today}

\begin{abstract}
We compute the vector and tensor contributions to the luminosity distance fluctuations in first order
perturbation theory, and we expand them in
spherical harmonics. This work presents the formalism with a first application to a 
stochastic background of primordial gravitational waves.
\end{abstract}


\maketitle

\section{Introduction}\label{s:intro}
The distance--redshift relation for far away objects plays an important role in cosmology. It has led Hubble,
or rather Lema\^\i tre~\cite{benz}, to discover the expansion of the Universe; and the distance--redshift 
relation to far away Supernovae type Ia is at the origin of last year's Nobel Prize in physics for the 
discovery of the accelerated expansion of the Universe~\cite{SN1a,SNnew}.

A next step that has been initiated recently considers the angular and redshift fluctuations
of the luminosity distance, which may also contain important information about our Universe~\cite{Hui,BDG,BDG.err,BDK}.
One important unsolved problem is the question how strongly the distance--redshift relation may be affected by
the fact that the actual Universe is not homogeneous and isotropic, but the matter distribution and also the geometry have fluctuations. To first order in perturbation theory these fluctuations can average out in the mean and are therefore expected to be small. 

However, it has been found that they are significantly larger than the naively expected value that would be of the order of the gravitational potential, namely, $\sim 10^{-5}$.   An analysis in first order gave fluctuations of the order of $10^{-3}$, hence 100 times larger than the naive estimate~\cite{BDG}. Recently, Ben-Dayan et al.~\cite{BenDayan}
have calculated  a second order contribution to the distance--redshift relation
of the order of $\sim 10^{-3}$. Evidently, if the second order term is as large as the first order, this 
means that perturbation theory cannot be trusted. On the other hand, fully nonlinear toy models, which have 
been studied in the past, always gave relatively small modifications of the luminosity distance if the size of
the fluctuations, spherical voids~\cite{voids} or parallel walls~\cite{walls}, is small compared to the Hubble scale.
Hence the problem remains open.

So far, the perturbative analyses of the distance--redshift relation have concentrated on scalar perturbations.
In this work, we want to study the contributions from vector and tensor perturbations on a 
Friedmann--Lema\^\i tre (FL) universe. This is interesting 
for several reasons. First of all, tensor perturbations are generically produced during inflation, and 
hence their contribution has to be added for completeness. Second, a passing gravitational wave from 
some arbitrary source does generate a tensor perturbation in the distance--redshift relation to any far 
away object and could, at least in principle, be detected in this way.
For single binary sources we have found that this effect is very small~\cite{Enea.master}; 
however, a stochastic background might lead to a detectable effect. Even though vector perturbations 
are usually not generated during inflation (and if they are they decay during the subsequent 
radiation dominated phase), they are relevant in many models with sources like, e.g., cosmic strings 
or primordial magnetic fields.  A third important motivation to study 
vector and tensor contributions comes from the fact that at second order in perturbation theory, scalars 
also generate vector and tensor perturbations~\cite{cris.vec,cris.ten}. In a complete second order treatment
these have to be included. With the formalism developed in this work, such an inclusion is straight forward.
We plan to report on the result of these second order contributions in a forthcoming paper~\cite{2nd}.
A similar program is carried {out} in Refs.~\cite{Schmidt1,Schmidt2}. There the authors discuss scalar,
vector, and tensor perturbations and split them into $E$ and $B$ modes. The treatment of these papers 
is, however, more adapted to describe distortions of surveys and weak lensing, but the convergence
calculated there is related to our distance fluctuations.

The paper is organized as follows. In the next section we discuss the luminosity-redshift relation perturbatively 
at first order. In Sec.~\ref{s:ten} we apply these results to tensor perturbations. We first derive the 
general first order expressions, which we then expand in spherical harmonics. We also give a numerical 
example for the gravitational wave background from inflation. In Sec.~\ref{s:vec} we treat vector perturbations
and in Sec.~\ref{s:con} we conclude. Some lengthy calculations and some details are deferred to four
Appendices.
\newline
{\bf Notation:} We use the metric signature $\left( -, +,+,+ \right)$. We denote the derivative w.r.t. the conformal time $\eta$ with a dot.

\section{The distance--redshift relation}\label{s:dist}
For an arbitrary geometry, defined through the metric $g$, a distance measure $D$ from a source moving with 
4-velocity $u_S=u(x_S)$ and an observer moving with 4-velocity $u_O=u(x_O)$ can be obtained as a solution of the Sachs focusing equation~\cite{sachs}:
\begin{equation}\label{sachsfoc}
\frac{d^2 D}{d\lambda^2} = -\left( \mathcal R + \left| \Sigma \right|^2 \right) D .
\end{equation}
Here $\lambda$ is the affine parameter of a lightlike geodesic $x^\mu(\la)$ from the source 
to the observer, $x^\mu(\la_S) = x^\mu_S$, $x^\mu(\la_O) = x^\mu_O$, and 
\begin{equation}
\mathcal R= \frac{1}{2} R_{\mu \nu} k^\mu k^\nu \qquad \text{with} \quad k^\mu =\frac{dx^\mu}{d\lambda},
\end{equation}
$k^\mu$ is the 4-velocity of the lightlike geodesic, and $\Sigma$ is the complex shear of 
the 'screen' defined below. The source and observer are
made out of baryons; hence we identify the 4-velocity field $u^\mu(x)$ with the
(baryonic and dark) matter velocity field.

Considering a thin light bundle with vertex at the source, the luminosity distance is given by
\begin{equation}
D_L = \left( 1 +   z \right) D\;, 
\end{equation}
where $z$ denotes the source redshift, defined by 
\begin{equation}\label{e:red}
1+  z = \frac{ \left. g_{\mu \nu} k^\mu u^\nu \right|_S}{ \left. g_{\mu \nu} k^\mu u^\nu \right|_O} = \frac{\omega_S}{\omega_O}.
\end{equation}

We are considering past light cones without caustics between the observer and source positions. This is well 
justified as we are treating small perturbations on a Friedmann background. See Ref.~\cite{EBD} for more details on the effect of caustics in the past light cone.

The complex shear of the light ray bundle, $\Sigma$, is defined as follows~\cite{ns}: Consider two spatial orthonormal vectors $e_1$ and $e_2$, which are normal to both the 4-velocity $u_O$ and $k$ at the observer position and which are parallel transported along $k$, such that $\nabla_k e_a = 0$ for $a = 1, 2$. The vectors $e_1$, $e_2$ are a basis of the so-called ’screen’. Note that we do not require that $u$ be parallel transported along $k$; hence $e_1$, $e_2$ are in general not normal to $u$ elsewhere than at the 
observer.  The complex shear is defined by
\begin{equation}
\Sigma = \frac{1}{2} g\left (\epsilon, \nabla_\epsilon k \right)  =  -\frac{1}{2} 
g\left ( \nabla_\epsilon\epsilon, k \right),\quad \text{with} \quad \epsilon = e_1 + i e_2 .
\end{equation}
We consider a light bundle with  vertex at the source\footnote{A light bundle with vertex at the observer yields the angular diameter distance, which is related by a factor $(1+z)^2$ to the luminosity distance which we determine here.}. This leads to the following initial conditions (more details are found in Appendix \ref{app:sachs}) for the Sachs focusing equation~(\ref{sachsfoc})
\begin{equation}\label{incond}
D \left( \lambda_S \right) =0, \qquad D' \left( \lambda_S \right)= \omega_S = - \left. g_{\mu \nu} k^\mu u^\mu \right|_S .
\end{equation}

In a perturbed FL metric the Sachs focusing equation~(\ref{sachsfoc}) reduces, at first order, to 
\begin{equation} \label{sachs1}
\frac{d^2 D}{d\lambda^2} = -\mathcal R  D.
\end{equation}
Since the complex scalar shear $\Sigma$ vanishes for a conformally flat spacetime, $|\Si|^2$ contributes
only at second order. To first order in $\mathcal R$, Eq.~(\ref{sachs1}) with initial conditions~(\ref{incond}) is solved by
\begin{eqnarray}\label{eq8}
\frac{D(\lambda_O)}{\om_S} &=&  \left( \lambda_O \! -\! \lambda_S \right)\! - \!\int_{\lambda_S}^{\lambda_O}\! \! d\lambda \!\!\int_{\lambda_S}^\lambda \! \!d\lambda' \mathcal R \left( \lambda\! -\! \lambda_S \right) \nonumber \\
&=&\left( \lambda_O \! -\! \lambda_S\right) \!- \!\int _{\lambda_S}^{\lambda_O}\!\! d\lambda ( \lambda\! -\! \lambda_S ) ( \lambda_O\! - \! \lambda ) \mathcal R ,
\end{eqnarray}
where we have used the  identity
\begin{equation}\label{e:idi}
\int_{\eta_S}^{\eta_O}\! d\eta \int_{\eta_S}^\eta \!  d\eta'  f\left( \eta' \right) =\int_{\eta_S}^{\eta_O} \! d\eta \left( \eta_O - \eta \right) f \left( \eta\right)
\end{equation}
for the second equal sign.

Of course, in a perturbed FL universe $\mathcal R$ is not first order; it also has a zeroth order contribution. 
But a perturbed FL universe is conformally related by the scale factor $a$ to  a perturbed Minkowski spacetime and lightlike 
geodesics are invariant under conformal transformations. Two conformally related metrics,
$$\tilde g_{\mu\nu} = a^2g_{\mu\nu}$$ 
 have the same lightlike geodesic curves, and only the affine parameter changes,  $d\tilde\la = a^2d\la$, 
 such that $ \tilde k^\mu = a^{-2}k^\mu$. Also the (normalized) matter 4-velocity changes, 
 $\tilde u^\mu = a^{-1}u^\mu$ so that the redshifts are related by
 \be \label{e:redOS}
 \tilde z + 1 = \frac{a_O}{a_S}\left( \delta z+1\right) \,, \quad \mbox{where  }~\frac{a_O}{a_S} \equiv \bar z +1
 \ee
 is the background redshift, i.e., the redshift in an unperturbed Friedmann--Lema\^\i tre  universe, and $\delta z$ is the source redshift according to definition~(\ref{e:red}) w.r.t. to the perturbed Minkowski metric $g$, while $\tilde z$ is the one w.r.t. to the perturbed FL metric $\tilde g$. We remark that $\tilde z$ is the true (observed) redshift.
In what follows we shall normalize the scale factor at the observer to one, $a_O=1$.
The distance $D$ is not affected by a conformal factor, so that the 
effect of the expansion on the distance simply leads to a rescaling~\cite{BDG}
\begin{equation}\label{eq11}
\tilde D_L = \left( 1 + \bar z \right) D_L \,.
\end{equation}

We now compute the luminosity distance in a perturbed Minkowski spacetime, $D_L$, and then relate it 
to the one in a FL spacetime, $\tilde D_L$, by the above rescaling.
Let $\left( 1, n^i \right)$ be the 0-order term of the lightlike velocity vector $k^\mu$ (in the 
nonexpanding Minkowski spacetime). The lightlike condition implies $\left| \bf n \right|^2 =  1$. 
We normalize the affine parameter $\lambda$ such that 
$\om_S=k_S^0 = 1$ at 0th order. To determine the redshift $\delta z$,
we have to solve the perturbed geodesic equation for $\mu =0$ only (in order to determine $k^0_O$ to first order), since the peculiar velocities are
already first order. The Christoffel symbols of Minkowski space vanish, so that the geodesic 
equation for $\mu=0$ to first order is simply
\begin{equation}\label{geodesic}
\frac{dk^0}{d \lambda} + \Gamma^0_{00} +2 \Gamma^0_{i0} n^i + \Gamma^0_{ij} n^i n^j=0.
\end{equation} 
We  normalize the affine parameter $\lambda$ such that $k^0_S=1$, and Eq.~(\ref{geodesic}) is solved by
\begin{equation} \label{geodsol}
k^0_O = 1\! -\! \int_{\lambda_S}^{\lambda_O}\hspace{-0.3cm} d\lambda \left(  \Gamma^0_{00} +2 \Gamma^0_{i0} n^i + \Gamma^0_{ij} n^i n^j  \right). 
\end{equation}
The geodesic equation~(\ref{geodesic}) will be useful also in order to express the distance $D$ in terms of the conformal time $\eta$ instead of the affine parameter $\lambda$. For this we use
\begin{equation}
k^0 = \frac{d \eta}{d\lambda}= 1 - \int_{\lambda_S}^{\lambda}\hspace{-0.3cm} d\lambda' \left(  \Gamma^0_{00} +2 \Gamma^0_{i0} n^i + \Gamma^0_{ij} n^i n^j  \right),
\end{equation}
which, in first order, leads to
\begin{equation}\label{eq13}
\lambda_O \!- \!\lambda_S \!=\! \eta_O \!-\! \eta_S \!+\! \!\int_{\eta_S}^{\eta_O}\hspace{-0.5cm} d \eta \!\!\int_{\eta_S}^\eta \hspace{-0.3cm}d \eta' \!\left(  \Gamma^0_{00} \!+\!2 \Gamma^0_{i0} n^i\! +\! \Gamma^0_{ij} n^i n^j  \right).
\end{equation}

The  conformal time and the background redshift are not observable. We want to write the distance as a function of the true (observed) redshift $\tilde z = \bar z + \de \tilde z$, where $\de \tilde z \equiv \left( 1 + \bar z \right) \delta z$ according to Eq.~(\ref{e:redOS}).
Following the approach presented in~\cite{BDG} we compute
\begin{eqnarray} \label{16}
\hspace{-0.5cm}\tilde D_L \left( \eta_S, \bf{n} \right) &=& \tilde D_L \left( \eta \left( \bar{z}\right), \bf n \right) \equiv \tilde D_L \left( \bar z , \bf n \right) \nonumber \\
&=& \tilde D_L \left( \tilde z , \bf n \right) - \left. \frac{d}{d\tilde z } \tilde D_L\left(\tilde z , \bf n \right)\right|_{\tilde z= \bar z} \delta \tilde z ,
\end{eqnarray}
with
\begin{eqnarray}\label{17}
\left. \frac{d}{d\tilde z } \tilde D_L \left( \tilde z, \bf n\right)\right|_{\tilde z= \bar z} &=& \frac{d}{d \bar z } \tilde D_L \left( \bar z, \bf n\right)  + \text{first order}\nonumber \\
&=&\left( 1 + \tilde z \right)^{-1} \tilde D_L +\mathcal{H}_S^{-1} + \text{first order}, \nonumber \\  
\text{where} \quad \mathcal{H}_S= \left. \frac{\dot a}{a} \right|_S. \hspace{-0.5cm}&&
\end{eqnarray}
In other words, we evaluate the distance at the true (observed) redshift $\tilde D_L\left( \tilde z, \bf n\right)$ by using Eqs.~(\ref{16}, \ref{17}) in order to relate $\tilde D_L ( \tilde z, \bn)$ to $\tilde D_L(\eta_S,\bn)$.

From Sec.~\ref{s:ten} on, to simplify the notation, we denote the true (observed) redshift with $z$ instead of $\tilde z$. We shall not use $\tilde z$ anymore.

\section{The distance--redshift relation from tensor perturbations}\label{s:ten}
We first consider a perturbed Minkowski metric with tensor perturbations only, defined by
\begin{equation}\label{metricten}
ds^2= -d\eta^2 + \left( \delta_{ij} + 2H_{ij}  \right) dx^i dx^j,
\end{equation}
where the tensor perturbations are divergence-free $ H^i_{j,i}=~0$, traceless $H^i_i=0$, symmetric $H_{ij}=H_{ji}$, and spatial $H_{\mu 0}=0$. By definition, a spin-2 perturbation is gauge--invariant. To use a notation consistent with the next section, we introduce the gauge invariant shear on the $\{t={\rm constant}\}$ hypersurfaces
$\sigma_{ij} = \dot H_{ij}$ (see,~e.g.,~\cite{RD}).

\subsection{The perturbation equations}
From the Ricci tensor calculated in Appendix~\ref{app:ten} we obtain
\begin{equation}\label{e:Rten}
\mathcal R = - \frac{1}{2} n^i n^j \Box H_{ij}, \qquad \text{where} \quad \Box = \partial^\mu \partial_\mu.
\end{equation}
Note that this is the Minkowski space d'Alembertian, without expansion.
The geodesic equation~(\ref{geodsol}) for $\mu=0$ leads to (see Appendix~\ref{app:ten} for details)
\begin{equation}
k^0_O = 1 - \int_{\lambda_S}^{\lambda_O} d\lambda \ \sigma_{ij} n^i n^j.
\end{equation}
We consider the 4-velocity $\left( u^\mu \right) = \left( 1 , {\bf 0 } \right)$ because the spin-2 perturbations can not source peculiar velocities at linear order, so that we obtain the redshift to first order
\begin{equation}
1+ \delta z = \frac{1 }{k^0_O  } =   1   +  \int_{\lambda_S}^{\lambda_O} d\lambda \ \sigma_{ij} n^i n^j .
\end{equation}
With Eqs.~(\ref{eq8}, \ref{eq11}, \ref{eq13}) we find the luminosity distance in a perturbed FL universe 
with $d\tilde s^2=a^2ds^2$, as a function of the background redshift
\begin{eqnarray}
\tilde D_L \left( \bar z, \bf n \right) &=& \left( 1 + \bar z \right) \left( \eta_O - \eta_S \right)  \nonumber\\
&& \hspace{-1.3cm} \times \left( 1 + \int_{\eta_S}^{\eta_O} d\eta \ \sigma_{ij} n^i n^j \right.   \nonumber \\
&& \hspace{-0.5cm} + \int_{\eta_S}^{\eta_O} d\eta\frac{\eta_O-\eta}{\eta_O-\eta_S}\sigma_{ij} n^i n^j \nonumber \\
&&  \hspace{-0.5cm}\left.- \int_{\eta_S}^{\eta_O} d\eta\frac{ \left( \eta - \eta_S \right) \left( \eta_O - \eta \right)}{\eta_O-\eta_S} \mathcal R \right).
\end{eqnarray}
We have again used (\ref{e:idi}) to reduce the double integral. We finally
express the luminosity distance in terms of the true, observed redshift $z$. Using Eqs.~(\ref{16},\ref{17}),  we obtain 
\begin{eqnarray}
\tilde D_L \left(  z, \bf n \right) &=& \left( 1 +  z \right) \left( \eta_O - \eta_S \right) \nonumber \\ 
&& \hspace{-1.3cm} \times \left( 1 - \frac{\mathcal H_S^{-1}}{\eta_O - \eta_S}  \int_{\eta_S}^{\eta_O} d\eta \  \sigma_{ij} n^i n^j  \right. \nonumber  \\
&& \hspace{-0.5cm}+ \int_{\eta_S}^{\eta_O} d\eta \frac{\eta_O-\eta}{\eta_O-\eta_S} \sigma_{ij} n^i n^j  \nonumber \\
&&  \hspace{-0.5cm} \left.- \int_{\eta_S}^{\eta_O} d\eta\frac{ \left( \eta - \eta_S \right) \left( \eta_O - \eta \right)}{\eta_O-\eta_S} \mathcal R \right) \! . \label{distten}
\end{eqnarray}
The origin of the different terms in the redshift--distance relation is as follows: the first line is the unperturbed
expression for the luminosity distance in a FL universe at the observed redshift $z$, the term on the second line derives from the redshift correction, the one on the third line from the relation between the conformal time $\eta$ and the affine parameter $\lambda$, and the one on the last line from the Sachs focusing equation. We can interpret this last term as a lensing effect.
The first two terms come from the perturbation of the redshift.

For a fluid with a vanishing anisotropic stress the redshift--distance relation becomes
\begin{eqnarray}
\tilde D_L \left(  z, \bf n \right) &=& \left( 1 +  z \right) \left( \eta_O - \eta_S \right)  \nonumber\\ 
&&\hspace{-2.05cm}\times \left( 1 - \frac{\mathcal H_S^{-1}}{\eta_O - \eta_S}  \int_{\eta_S}^{\eta_O} d\eta \  \sigma_{ij} n^i n^j  \right. \nonumber  \\
&&\hspace{-1.2cm}+ \int_{\eta_S}^{\eta_O} d\eta\frac{\eta_O-\eta}{\eta_O-\eta_S} \sigma_{ij} n^i n^j \nonumber \\
&& \hspace{-1.2cm}  \left.+ \int_{\eta_S}^{\eta_O}  \hspace{-0.2cm}d\eta \frac{ \left( \eta - \eta_S \right) \left( \eta_O - \eta \right)}{\eta_O-\eta_S} \mathcal{H} \ n^i n^j \sigma_{ij} \right), \label{disttenideal}
\end{eqnarray}
where we used the Einstein equation~\cite{RD}
\begin{equation}\label{e:Ei}
\ddot H_{ij} + 2 \mathcal{H} \dot H_{ij}  -\nabla^2 H_{ij} =0,
\end{equation}
and Eq.~(\ref{e:Rten}) to replace $\mathcal R$. 
If the cosmic fluid is not ideal, but  has anisotropic stresses, these add to the right-hand side of Eq.~(\ref{e:Ei}) (see~\cite{RD}) and correspondingly to the last line in Eq.~(\ref{disttenideal}), the lensing term.

\subsection{Spherical harmonic analysis}
We want to determine the power spectrum of the luminosity distance. In the unperturbed FL background the luminosity distance to the redshift $z$ is given by
\begin{equation}
\bar D_L \left(  z \right) = \left( 1 +  z \right) \left( \eta_O - \eta_S \right).
\end{equation}
We define the relative difference in the luminosity distance as
\bea
\Delta_L \left( z, \bf n \right) &=& \frac{\tilde D_L \left( z, \bf n \right) - \bar D_L\left( z \right)}{\bar D_L\left( z \right)}
\nonumber \\
&=&  \frac{1}{\eta_O - \eta_S}  \int_{\eta_S}^{\eta_O} \!\!\! d\eta \Big[-\HH_S^{-1} +( \eta_O - \eta) + \nonumber  \\
&&+ \left( \eta - \eta_S \right) \left( \eta_O - \eta \right)\HH\Big] \sigma_{ij} n^i n^j  
\,. \label{e:29}
\eea
Note that we evaluate the unperturbed distance at the true, observable redshift.

We are interested in the angular power spectrum of this observable, $c_\ell \left( z, z' \right) $, which depends
on the redshift of the two sources and is defined by the two point correlation function
\begin{eqnarray} 
 &&\hspace{-1.5cm} \langle\Delta_L \left( z, \bf n\right) \Delta_L\left( z', \bf n' \right) \rangle  \nonumber \\
 &=&  \frac{1}{4 \pi} \sum_\ell \left( 2 \ell +1 \right)  c_\ell(z,z') P_\ell \left( \bf n \cdot \bf n' \right) . \label{29b}
\end{eqnarray} 

In the distance--redshift relation~(\ref{distten}) [and, in particular, in Eq.~(\ref{disttenideal}) for an ideal fluid] 
we have several times the term $n^i n^j \sigma_{ij}\left( \eta, \bf x \left( \eta \right) \right) $ where 
${\bf x} \left( \eta \right) = {\bf x}_O - {\bf n} \left( \eta_O - \eta \right)$. In terms of its Fourier transform this is
\begin{equation}
n^i n^j \sigma_{ij}\left( \eta, \bf x \left( \eta \right) \right) \!=\! \!\int \!\frac{d^3k}{\left( 2 \pi \right)^3} \hat \sigma_{ij} \left( \eta, \bf k \right)n^i n^j e^{-i \bf k \cdot \bf x(\eta)}.
\end{equation}
Without loss of generality we choose ${\bf x}_O ={\bf 0} $. Setting
\begin{equation}
{\bf k} = {\bf \hat k } \left| {\bf k} \right| = {\bf \hat k}  k, \ \ \mu = { \bf \hat k} \cdot {\bf n} ,   \  \  \Delta\eta = \eta_O -\eta, 
\end{equation}
we  obtain
\begin{equation}
n^i n^j \sigma_{ij}\left( \eta, \bf x \left( \eta \right) \right)  = \int  \frac{d^3k}{\left( 2 \pi \right)^3} \  \hat \sigma_{ij}\left( \eta,  \bf k \right) n^i n^j \ e^{i \mu  k \Delta\eta }.
\end{equation}
Writing the exponential in terms of spherical Bessel functions
\begin{equation}
 e^{i \mu k \Delta\eta  } = \sum_{ l=0}^{\infty} \left( 2  \ell + 1 \right) i^{ \ell } \  j_{ \ell } \left( k\Delta\eta  \right) P_{ \ell } \left( \mu \right)  ,
\end{equation}
we find
\begin{equation} \label{34}
n^i n^j \sigma_{ij} = \int    \frac{d^3k}{\left( 2 \pi \right)^3}   n^i n^j \hat \sigma_{ij}  \sum_{ \ell=0}^{\infty} \left( 2  \ell + 1 \right) i^{\ell } j_{\ell } \left( k\Delta\eta  \right) P_{ \ell } \left( \mu \right).
\end{equation}
With respect to a helicity basis  in Fourier space 
\be\label{e:hel}
\left\{ \bf e^{( + )},e^{( - )}, {\bf \hat k} \right\} \;,\quad 
{\bf e}^{( \pm)} = \frac{1}{\sqrt{2}}\left({\bf e}_1\pm i {\bf e}_2\right)\,, \ee
such that
\begin{equation} 
{ \bf e^{\left( \pm \right) } \cdot  n}= \sqrt{\frac{1 - \mu ^2}{2}} e^{\pm i \phi},
\end{equation}
we have
\begin{equation}
\hat \sigma_{ij} \ {\bf e}^i \otimes {\bf e}^j = \hat \sigma^{+ } \ {\bf e}^{\left( + \right)} \otimes {\bf e}^{\left( + \right)}   + 
\hat \sigma^{- } \ {\bf e}^{\left( - \right)} \otimes {\bf e}^{\left( - \right)}.
\end{equation}
We introduce the spherical harmonics with respect to some arbitrary $z$ direction given by a unit vector $\bf e$ as
$Y_{\ell m}({\bf n}, {\bf e})$, since we shall use them w.r.t. different $z$ axes. The addition theorem of spherical harmonics
is
\begin{equation}\label{addtheom}
P_{ \ell } \left( \mu \right) = \frac{4 \pi }{2  \ell + 1} \sum_{ m} Y^*_{ \ell  m} \left( \bf \hat k ,\bf e \right) 
Y_{ \ell  m} \left( \bf n, \bf e \right).
\end{equation}
Using the following spherical harmonics definition
\begin{equation}
Y_{2,\pm 2} \!\left( \bf n ,\! \bf \hat  k \right)\!=\!   \sqrt{\!\frac{15}{8\pi}} \sin^2 \!\theta   e^{\pm 2i \phi} \!=\! 
\sqrt{\!\frac{15}{2 \pi }}\frac{1 \!-\! \mu^2}{2} e^{\pm 2i \phi},
\end{equation}
we can rewrite Eq.~(\ref{34}) as
\begin{eqnarray}
n^i n^j \sigma_{ij} &=&\int \frac{d^3k}{\left( 2 \pi \right)^3}\left(\hat \sigma^+ Y_{22} \left( \bf n , \hat k \right)+ \hat \sigma^- Y_{2-2} \left( \bf n , \hat k \right)  \right) \nonumber \\
&& \hspace{-1cm}\times  \sqrt{\frac{2 \pi }{15}}   \sum_{ \ell,  m} 4 \pi  i^{\ell} j_{\ell } \left( k \Delta\eta \right) Y^*_{ \ell  m}  \left( \bf \hat k , \bf e \right) Y_{\ell m} \left( \bf n, \bf e \right) . 
\end{eqnarray}
We now introduce the initial tensor power spectrum $P_H\left( k \right)$ through
\begin{eqnarray}
&&\hspace{-0.7cm} \langle\hat H^{\pm}\left( \eta_S, \bf  k \right)  \hat  H^{\pm *}\left( \eta_{S'}, \bf k' \right) \rangle   \nonumber \\
&&  = \left( 2 \pi \right)^3  \delta^{\left( 3 \right)} \left( {\bf k }- {\bf k' } \right) P_{H} \left( k \right) T_k \left( \eta_S \right) T_k \left( \eta_{S'} \right),
\end{eqnarray}
where $T_k \left( \eta \right)$ is the transfer function with the initial condition $T_k(\eta) \ra_{(k\eta\ra 0)} 1$. 
The $\delta^{(3)}(\bf k - \bf k')$ is a consequence of stochastic homogeneity. For the shear we then have
\begin{eqnarray}
&&\hspace{-0.7cm} \langle\hat \sigma^{\pm}\left( \eta_S, \bf  k \right)  \hat  \sigma^{\pm *}\left( \eta_{S'}, \bf k' \right)\rangle 
\nonumber \\
&& = \left( 2 \pi \right)^3  \delta^{( 3 )}\! \left( {\bf k }- {\bf k' } \right) P_{H} \left( k \right) \dot T_k \left( \eta_S \right) \dot T_k \left( \eta_{S'} \right). 
\end{eqnarray}
Next, we express the terms in the distance--redshift relation with the help of the power spectrum of the integrand,
\begin{equation}
\langle n^i n^j  \sigma_{ij} n'^l n'^k  \sigma_{lk}\rangle = \frac{1}{4 \pi } \sum_\ell \left( 2\ell +1 \right) \bar c_\ell(\eta,\eta')
 P_\ell \left( \bf n \cdot n' \right).
\end{equation}
A lengthy but straight forward calculation yields~\cite{DK,RD}
\begin{equation}
\bar c_\ell \!= \!  \frac{1}{ \pi} \frac{\left( \ell \!+\!2 \right)!}{\left( \ell \!-\!2 \right)!} \!\int\!\! dk \, k^2 P_H( k ) \dot T_k( \eta) \dot T_k(\eta') \frac{j_{ \ell}( k\Delta\eta) }{( k \Delta\eta)^2}   \frac{j_{ \ell }( k \Delta\eta') }{( k \Delta\eta')^2}.
\end{equation}

Using the Limber approximation (see Appendix~\ref{app:lim}) for the time integrals in Eq.~(\ref{disttenideal}), 
and, in particular, Eqs.~(\ref{limb3}),~(\ref{limb4}) and~(\ref{limb5}),
 we can simplify the time integrals, and we find the coefficients (under the ideal fluid assumption)
\begin{eqnarray}
c_\ell(z,z') &\simeq& \frac{1}{\pi } \frac{\left( \ell +2 \right)!}{\left( \ell -2 \right)!} \frac{I_\ell^2}{\ell^4}
          \frac{1}{\Delta \eta_S}\frac{1}{\Delta \eta_{S'}}  \nonumber \\
&&\hspace{-2.2cm} \times \! \!\int_{k^*}^\infty \hspace{-0.32cm}dk \, P_H( k) \dot T^2(\eta_{\ell ,k}) \left( A +B \mathcal H (\eta_{\ell, k})+ C  \mathcal H^2(\eta_{\ell, k})\right)\! , \hspace{0.5cm}    \label{cltensor}
\end{eqnarray}
where we have introduced
\begin{eqnarray}
 \Delta \eta_{S}  =  \eta_O - \eta_{S}, &\qquad &  \Delta \eta_{S'}  =  \eta_O - \eta_{S'} , \\
 \eta_{\ell, k} = \eta_O - \frac{\ell}{k}, &\qquad& I_\ell^2 = \frac{1.58}{\ell},  \\
 k^*= \text{max}&& \hspace{-1.1 cm}\left\{ \frac{\ell}{\Delta\eta_S},\frac{\ell}{\Delta\eta_{S'}} \right\},
 \end{eqnarray}
and
\begin{eqnarray}
A &=& \mathcal H_S^{-1}\mathcal H_{S'}^{-1} + \frac{\ell^2}{k^2} - \frac{\ell}{k} \left( \mathcal H_S^{-1} + \mathcal H_{S'}^{-1} \right) , \\
B&=& - \frac{\ell}{k} \left( \mathcal H_S^{-1} \Delta \eta_{S'}  + \mathcal H_{S'}^{-1} \Delta \eta_{S}  \right)  
     - 2 \frac{\ell^3}{k^3} \nonumber \\
&& + \frac{\ell^2}{k^2} \left( \Delta \eta_S + \Delta \eta_{S'} + \mathcal H_S^{-1} +\mathcal H_{S'}^{-1} \right), \\
C&=& \frac{\ell^2}{k^2 }\Delta \eta_S \Delta \eta_{S'} - \frac{\ell^3}{k^3} \left( \Delta \eta_S + \Delta \eta_{S'} \right) + \frac{\ell^4}{k^4}.
\end{eqnarray}
More details can be found in Appendix~\ref{app:lim}.

\subsection{Application}
As an example, we consider a flat primordial tensor power spectrum as expected from inflation 
$P_H \left( k \right) = \alpha /k^3$.
If $r$ denotes the tensor to scalar ratio, the scalar amplitude as measured  by the Wilkinson Microwave Anisotropy Probe experiment~\cite{WMAP} yields $\alpha\simeq r\times 10^{-9}$, such that the tensor power spectrum becomes
\begin{equation}
P_H \left( k \right)  \simeq \frac{r }{k^3} \ 10^{-9}.
\end{equation}
Considering an ideal fluid, the transfer function $T_k \left( \eta \right)$ is the solution of the differential 
equation~(\ref{e:Ei}),
\begin{equation}\label{e:ten}
\ddot T_k \left( \eta \right) + 2 \mathcal H \dot T_k \left( \eta \right)  + k^2 T_k \left( \eta \right) =0,
\end{equation}
with initial condition $T_k(\eta_{\rm in}) =1$ and  $\dot T_k(\eta_{\rm in}) =0$ for $k\eta_{\rm in}\ll 1$.
In a matter (or radiation) dominated universe this differential equation can be solved analytically  in terms of Bessel functions. The growing (not decaying) mode is given by
\begin{equation}
T_k \left( \eta \right) = \left( k \eta \right)^{1/2-q} Y_{1/2-q}\left( k \eta \right) , \quad \text{where} \quad a \propto \eta^q, 
\end{equation} 
and $Y_\nu$ is the Bessel function of the second kind of order $\nu$. At late times, when the cosmological constant dominates, we cannot write the scale factor $a \left( \eta \right)$ as a power law and we have no analytic solution
to~(\ref{e:ten}). To determine the $c_\ell$ coefficients, we have solved the differential Eq.~(\ref{e:ten}) numerically. 

\begin{figure}[ht]
\includegraphics[width=8.5cm]{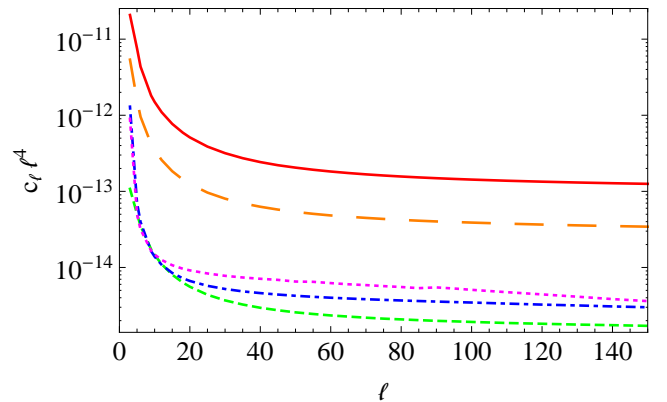}
\caption{We show the tensor power spectrum rescaled by $\ell^4$ for the fluctuations in the luminosity distance for different values of the source redshift ($z= 0.5$, dotted pink line; $z=1$, dot-dashed blue line;  $z=2$, dashed green line;  $z=3$, long-dashed orange line;  $z=4$, solid red line). In the figure we have set $r=1$.\label{f:spec1}}
\end{figure}

The resulting power spectrum $c_\ell(z,z)$ for different source redshifts is shown in Fig.~\ref{f:spec1}.

Clearly, for sufficiently large $\ell$, $c_\ell(z,z) \propto \ell^{-4}$. 
The simplest way to understand this scaling is to note that once a mode enters the horizon, the tensor fluctuations scale like $ \int\si d\eta \sim H\propto a_k/a \propto (k\eta)^{-q}$, where $a_k=a(\eta=1/k)$ denotes the value of the scale factor at horizon entry. For modes that enter during the radiation era $q=1$, while for modes
that enter during the matter era $q=2$. Hence $\int \si d\eta \propto H \propto
H_{\rm in}/k^q$  is acquiring a factor $k^{-q}$ with respect to the scale invariant initial spectrum. This leads to a red 
spectrum, $k^3\left(\int \si\right)^2 \propto k^{-2q}$ and $\ell^2 c_\ell(z,z) \propto \ell^{-2q}$.
This spectrum turns from $c_\ell \propto \ell^{-4}$ for scales that enter the horizon in the radiation era to
$c_\ell \propto \ell^{-6}$ for scales that enter the horizon in the matter era. For $z=4$ this happens roughly at
$\ell \sim 20$. Of course, the transition is quite gradual.

Comparing Fig.~\ref{f:spec1} with the results from scalar perturbations~\cite{BDG}, we see first that the tensor contribution is much smaller, nearly 8 orders of magnitude. We obtain $\ell^4c_\ell(z) \sim 5\times10^{-13}$ for $z=4$ and $\ell\gsim 40$
while scalar perturbations yield $\ell^2c_\ell(z) \sim 10^{-5}$   for $z=4$ and $\ell\gsim 100$. Furthermore,
despite also being proportional to the lensing term, it scales differently with $\ell$. This comes from the fact that the scalar
lensing term is determined by the spectrum of $k^2\Psi$, where $\Psi$ is the scale invariant Bardeen potential, while for
scales that enter during radiation dominated expansion, $\si_{ij}n^in^j$ is suppressed by a factor of $1/k$.

\begin{figure}[ht]
\includegraphics[clip=true,width=8.5cm]{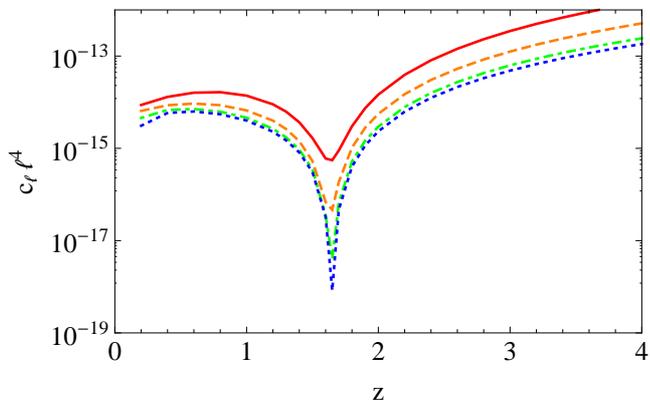}
\caption{We show the tensor power spectrum rescaled by $\ell^4$ for the fluctuations in the luminosity distance as
a function of the source redshift  for different values $\ell$ ($\ell=60$, dotted blue line;  $\ell=40$, dot-dashed green line;  $\ell=20$, dashed orange line;  $\ell=10$, solid red line). Also here $r=1$.\label{f:spec2}}
\end{figure}

Interestingly the tensor signal is not monotonic in redshift up to $z\simeq 2$.  It has a sharp minimum at $z \simeq 1.65$.  
To illustrate this, we also plot $c_\ell(z,z)$ as a function of the
source redshift for different values of $\ell$ in Fig.~\ref{f:spec2}.

\begin{figure}[ht]
\includegraphics[clip=true,width=8.5cm]{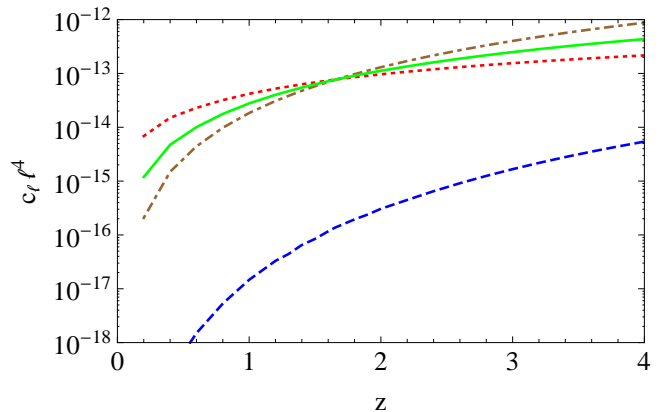}
\caption{We show the contributions from the different terms of Eq.~(\ref{e:29}) to the tensor power spectrum rescaled by $\ell^4$ for the 
fluctuations in the luminosity distance as
a function of the source redshift  for $\ell=40$. (First term: dotted red line; second term: dot-dashed brown line; third term: dashed blue line;  correlator of the first and second term: solid green line). The third term is always subdominant. 
We plot the correlator between the first and the second term (solid line) with the opposite sign since it is negative. 
The second and the first terms have opposite signs, and they cross at $z_c\simeq 1.65$.\label{f:terms}}
\end{figure}

The signal drops to $0$ at $z_c=1.65$. This comes from the fact that it is dominated by two terms with opposite sign. To see this, we also show the contributions from the three terms in the square bracket of~(\ref{e:29}) individually in Fig.~\ref{f:terms}. 

If the source redshift is small, $\eta_S\sim \eta_O$, the  first term $\propto -\HH_S^{-1} \sim -\eta_S/2$ dominates, 
while for large redshifts, $\eta_S\ll\eta_O$, the second term $\propto (\eta_O-\eta)$ dominates.
If $\si_{ij}n^in^j$ has a definite sign, the result inherits this sign for small redshifts and the opposite sign for large 
redshifts. The sign changes happens around $\eta_S=\eta_O/2$ corresponding
to a redshift  $z_c\sim 3$. This is not expected to be very precise; in particular, we have neglected the time dependence of the transfer function in this argument. The more precise numerical evaluation
gives $z_c \simeq 1.65$. Interestingly this redshift is close to the maximum of the angular diameter distance 
$D_A(z)=(\eta_O-\eta_S)/(1+z)$.

The results shown in Figs.~\ref{f:spec1} to \ref{f:terms} have been calculated with the following cosmological 
parameters: $h=0.7$, $H_O^{-1} =2997.9 h^{-1} \text{Mpc}$, $\Omega_m h^2=0.13$, $\Omega_r h^2 = 
4.17 \times 10^{-5}$, and $\Omega_\Lambda= 1 - \Omega_m - \Omega_r$.

\section{The distance--redshift relation from vector perturbations}\label{s:vec}
We now consider vector perturbations. As for tensor perturbations, we can divide out the cosmic expansion for lightlike geodesics. Hence we can consider Minkowski space with purely vector perturbations. The metric is then given by
\begin{equation}\label{metricvec}
ds^2\!= -d\eta^2 \!- 2 B_i dx^i d\eta + \left( \delta_{ij} + H_{i,j} + H_{j,i} \right) dx^i dx^j,
\end{equation}
where the perturbations are divergence-free, $B^i_{,i}=~H^i_{,i}=~0$. Using $B_{ij} = B_{\left( i,j \right) }$ and $ H_{ij}= H_{\left( i,j \right)} $, where $(~)$ denotes symmetrization, the shear on the constant time hypersurface
is given by~\cite{RD}  $\sigma_{ij} =  B_{i j} + \dot H_{ij}$ or in  3-vector notation 
$\sigma_i = B_i + \dot H_i $, and  $\sigma_{ij} =\sigma_{(i,j)}$. This quantity is gauge invariant~\cite{RD}.

\subsection{The perturbation equations}
With the Ricci tensor calculated in Appendix \ref{app:vec} we obtain for vector perturbations
\begin{equation}\label{e:Rvec}
\mathcal R = \frac{1}{2} \left( \nabla^2 \left( \sigma_i n^i \right) + \dot \sigma_{ij} n^i n^j \right).
\end{equation}
To determine the redshift  we first evaluate the geodesic solution~(\ref{geodsol}) with the Christoffel symbols derived in Appendix~\ref{app:vec}, 
\begin{equation}
k^0_O = 1 - \int_{\lambda_S}^{\lambda_O} d\lambda \ \sigma_{ij} n^i n^j.
\end{equation}
Vector perturbations can have a nonvanishing peculiar velocity term. 
We define the observer 4-velocity $(u^\mu) = \left( 1 , B^i + v^i \right)$.  The peculiar velocity $v^i$ defined in
this way is gauge invariant. It is the vorticity of the matter flow. We now obtain

\begin{equation}
1+ \delta z = \frac{-1 + n_i  v_S^i}{-k^0_O + n_i v_O^i } =   1+ n_i \left( v_O^i - v_S^i\right)  +    \int_{\lambda_S}^{\lambda_O} \hspace{-0.2cm}d\lambda \ \sigma_{ij} n^i n^j .
\end{equation}
After a short calculation, using the results of Sec.~\ref{s:dist}, we find the luminosity distance
\begin{eqnarray}
\tilde D_L \left( \bar z, \bf n \right) &=& \left( 1 + \bar z \right) \left( \eta_O - \eta_S \right)  \nonumber \\
&& \hspace{-1cm}\times \left( 1 +n_i v^i_O - 2 n_i v_S^i + \int_{\eta_S}^{\eta_O} \hspace{-0.31cm}d\eta \ \sigma_{ij} n^i n^j \right.  \nonumber \\
&&  +\frac{1}{\eta_O-\eta_S} \int_{\eta_S}^{\eta_O}\hspace{-0.31cm} d\eta (\eta_O-\eta) \sigma_{ij} n^i n^j \nonumber \\
&& \left.- \int_{\eta_S}^{\eta_O} \hspace{-0.31cm}d\eta  \frac{ \left( \eta - \eta_S \right) \left( \eta_O - \eta \right)}{\eta_O-\eta_S} \mathcal R \right).\label{22}
\end{eqnarray}
Since we are interested in expressing the luminosity distance as a function of the true redshift, we have to evaluate Eq.~(\ref{22}) at $z$ and subtract the correction term defined through Eqs.~(\ref{16}) and~(\ref{17}),
\begin{eqnarray}\label{57b}
\tilde D_L \left(  z, \bf n \right) &=& \left( 1 +  z \right) \left( \eta_O - \eta_S \right) \nonumber \\
&& \hspace{-1.52cm}\times \left( 1 - \frac{\mathcal H_S^{-1}}{\eta_O - \eta_S} n_i v^i_O -  n_i v_S^i \left( 1 -  \frac{\mathcal H_S^{-1}}{\eta_O - \eta_S}\right) \right. \nonumber  \\
&&  \hspace{-0.51cm}  - \frac{\mathcal H_S^{-1}}{\eta_O - \eta_S}  \int_{\eta_S}^{\eta_O}\hspace{-0.31cm} d\eta \ \sigma_{ij} n^i n^j   \nonumber \\
&& \hspace{-0.51cm}+\frac{1}{\eta_O-\eta_S} \int_{\eta_S}^{\eta_O} \hspace{-0.31cm}d\eta (\eta_O-\eta)
    \sigma_{ij} n^i n^j \nonumber \\
&&  \hspace{-0.51cm}\left.- \int_{\eta_S}^{\eta_O} \hspace{-0.31cm}d\eta  \frac{ \left( \eta - \eta_S \right) \left( \eta_O - \eta \right)}{\eta_O-\eta_S} \mathcal R \right), \label{distvec}\\
\Delta_L \left( z, \bf n \right) &=& - \frac{\mathcal H_S^{-1}}{\eta_O - \eta_S} n_i v^i_O -  n_i v_S^i \left( 1 -  \frac{\mathcal H_S^{-1}}{\eta_O - \eta_S}\right) \nonumber  \\
&&  \hspace{-0.51cm}  - \frac{\mathcal H_S^{-1}}{\eta_O - \eta_S}  \int_{\eta_S}^{\eta_O}\hspace{-0.31cm} d\eta \ \sigma_{ij} n^i n^j   \nonumber \\
&& \hspace{-0.51cm}+\frac{1}{\eta_O-\eta_S} \int_{\eta_S}^{\eta_O} \hspace{-0.31cm}d\eta(\eta_O-\eta)
 \sigma_{ij} n^i n^j \nonumber \\
&&  \hspace{-0.51cm} - \int_{\eta_S}^{\eta_O} \hspace{-0.31cm}d\eta  \frac{ \left( \eta - \eta_S \right) \left( \eta_O - \eta \right)}{\eta_O-\eta_S} \mathcal R . \label{distvec2}
\end{eqnarray}

This expression depends only on the gauge-invariant quantities $v^i$ and $\sigma_{ij}$ as it should. 
We note also that we did not assume any gravitational theory yet. Indeed the procedure used so far is completely geometrical. If one is interested in general relativity (GR), then the two gauge-invariant variables $v^i$ and $\sigma_{ij}$ are not independent but related via Einstein's equations~\cite{RD},
\begin{equation}\label{vecEinstein}
\nabla^2 \sigma_i = -16 \pi G a^2 v_i \left( \bar \rho + \bar p\right),
\end{equation}
where $\bar \rho$ and $\bar p$ are the background density and pressure, respectively.

\subsection{Spherical harmonic analysis}
As for the tensor perturbations, we are interested in the term $n^i n^j \sigma_{ij}$. The main difference is that in the vector case we have $\sigma_{ij}= \sigma_{\left( i,j \right)}$ in real space and $\hat \sigma_{ij} = -i k_{\left( i \right.} \hat \sigma_{\left. j \right)}$ in Fourier space. This leads to
\begin{equation}
n^l n^j \hat  \sigma_{lj} e^{i \mu k \Delta\eta } = - k n^j \hat \sigma_j \frac{\partial \ e^{i \mu k \Delta\eta } }{\partial \left( k\Delta\eta\right)} .
\end{equation}
With this we can write Eq.~(\ref{34}) as
\begin{equation}
n^i n^j \sigma_{ij} \!=\! - \int    \frac{d^3k}{\left( 2 \pi \right)^3}  k  \ n^j \hat \sigma_{j} \! \sum_{ \ell=0}^{\infty} \left( 2 \ell + 1 \right) i^{\ell } \  j'_{\ell } \! \left( k\Delta\eta \right) P_{ \ell }\! \left( \mu \right).
\end{equation}
With the helicity basis defined in Sec.~\ref{s:ten}, the addition theorem of the spherical 
harmonics~(\ref{addtheom}), and
\begin{equation}
Y_{1,\pm 1} \!\left( \bf n , \bf \hat  k \right)\!= \mp  \sqrt{\frac{3}{8\pi}} \sin \theta \  e^{\pm i \phi} = \mp \sqrt{\frac{3}{4 \pi }} \sqrt{\frac{1 - \mu^2}{2}} e^{\pm i \phi},
\end{equation}
we obtain
\begin{eqnarray} \label{46}
n^i n^j \sigma_{ij} &=&\int   \! \frac{d^3k}{\left( 2 \pi \right)^3}   k \left( Y_{1,1}\! \left( \bf n , \bf \hat  k \right) \hat \sigma^+ \!- \! Y_{1,-1}\!\left( \bf n , \bf  \hat  k \right) \hat \sigma^- \right)  \nonumber \\
&&\hspace{-0.5cm}\times  4 \pi \sqrt{\frac{4 \pi}{3} }  \sum_{ \ell,m}  i^{\ell } \  j'_{\ell } \left( kr \right) Y^*_{ \ell  m}( {\bf \hat  k}, {\bf e} ) 
Y_{ \ell  m}( {\bf n}, {\bf e}).
\end{eqnarray}

If we assume that vector perturbations have been generated at some time in the past, we can define the vector power spectrum as for the tensor case as
\begin{eqnarray}
&& \hspace{-0.8cm} \langle\hat \sigma^{\pm}\left( \eta_S, \bf  k \right)  \hat  \sigma^{\pm*}\left( \eta_{S'}, \bf k' \right)\rangle \nonumber\\
&& \quad=  \left( 2 \pi \right)^3  \delta^{\left( 3 \right)} \left( {\bf k }- {\bf k' } \right) P_{\sigma} \left( k \right) T_k \left( \eta_S \right) T_k \left( \eta_{S'} \right). 
\end{eqnarray} 
If we do not want to consider the case of early generation, we simply have to replace 
$P_{\si}( k )T_k(\eta_S)T_k( \eta_{S'} )$ by a time-dependent power spectrum, $P_\si(k,\eta_S,\eta_{S'})$. The model
under consideration (e.g., cosmic strings) then has to be used to determine this time-dependent power spectrum.
If, however, vector perturbations evolve freely, we can then compute the shear power spectrum as for tensors,
\begin{equation}\label{48}
\langle n^i n^j  \sigma_{ij} n'^l n'^k  \sigma_{lk}\rangle = \frac{1}{4 \pi } \sum_\ell \left( 2\ell +1 \right) \bar c_\ell(\eta,\eta') P_\ell \left( \bf n \cdot n' \right),
\end{equation}
finding (see Appendix~\ref{app:veccl}) 
\begin{eqnarray} 
\bar c_\ell&=&  \frac{2\ell \left(\ell +1 \right)}{ \pi \left( 2\ell +1 \right)^2} 
\int dk \ k^4 P_\sigma \left( k \right) T_k \left( \eta \right)  T_k \left( \eta' \right)   \nonumber \\
&&\hspace{-0.4cm}\times \left[ j'_{\ell-1} \left( kr \right) j'_{\ell-1} \left( kr' \right) +j'_{\ell+1} \left( kr \right) j'_{\ell+1} \left( kr' \right)  \right].  \label{clvectorgen}
\end{eqnarray}
As mentioned above,
in  more realistic scenarios, where vector perturbations are generated, e.g., via anisotropic stresses from topological defects or by second order perturbations, we obtain a shear power spectrum of the form $P_\sigma \left( k,\eta,\eta' \right)$, 
which cannot be factorized into a random initial spectrum and a deterministic transfer function.

In Appendix~\ref{app:veccl} we nevertheless, for sake of completeness, continue with expression~(\ref{clvectorgen})
to derive the vector angular power spectrum for the luminosity distance fluctuations. We do not repeat the 
lengthy, complicated, and not very illuminating formulas here.

\section{Conclusions and outlook}\label{s:con}
In this paper we have calculated the angular power spectrum of the linear vector and tensor fluctuations in the 
distance--redshift relation.  For vector perturbations we have simply derived the formulas and for tensor fluctuations we have applied them to an initial spectrum of fluctuations from inflation. It is interesting to see that the tensor-distance fluctuation spectrum is not simply suppressed by a factor $r$ as one might naively expect, but by  about $8$
orders of magnitude more. The reason for this is mainly that tensor fluctuations decay once they enter the 
horizon, while, on the contrary, scalar perturbations start growing.
We therefore expect that the tensor signal generated from scalar perturbations at second order dominates 
over the small first order signal. The calculations of these second order contributions are left to a future project~\cite{2nd}.

We have also found that the tensor signal is not monotonically increasing with redshift as we would expect it from a pure
lensing signal. This is due to the fact that the total signal is the sum of a  redshift part, $\propto - \de z/(z+1)$,
and a  lensing part. At redshift $z_c\simeq 1.65$, which is close to the redshift where the angular diameter distance $D_A=(\eta_O-\eta_S)/(1+z)$ has a maximum, these terms cancel, and at higher redshifts the redshift-term dominates.

\acknowledgments
This work is supported by the Swiss National Science Foundation.

\appendix

\section{Sachs focusing equation} \label{app:sachs}
Much of this work is based on the Sachs focusing equation~\cite{sachs}. It has been shown~\cite{Perlick} that the distance can be defined as
\begin{equation} \label{app1:1}
D = \sqrt{ \left| \det \mathcal D \right| },
\end{equation}
where $\mathcal D$ is the Jacobi matrix that satisfies the differential equation
\begin{equation}
\frac{d^2{ \mathcal D}}{d\lambda^2} = \left( \begin{array} {cc} - \mathcal R - \text {Re}  \left( F \right)  & -  \text {Im}  \left( F \right)  \\   \text {Im}  \left( F \right)  & - \mathcal R +  \text {Re}  \left( F \right)  \end{array} \right) \mathcal D, 
\end{equation}
with $F= \frac{1}{2} R_{\alpha \mu \beta \nu} \bar \epsilon^\alpha \bar \epsilon^\beta k^\mu k^\nu$.
The determinant of the Jacobi matrix $\mathcal D$ describes the area of the thin light beam and its square root is therefore a distance,~(\ref{app1:1}), if the affine parameter $\lambda$ is normalized such that $\omega_S=1$. In general one can use the Sachs focusing equation also with a different affine parameter normalization. Indeed from the distance definition
\begin{equation}
D = \sqrt{\frac{d A_O}{d \Omega_S}}
\end{equation}
and the solid angle aberration~\cite{sachs} 
\begin{equation}
\frac{d \tilde \Omega}{d \Omega} = \left( \frac{   k_\mu u^\mu  }{k_\mu \tilde u^\mu} \right)^2= \frac{\omega^2}{\tilde \omega^2},
\end{equation}
we find, setting $\omega =1 $, 
\begin{equation}
\tilde D =  \sqrt{\frac{d A_O}{d \tilde \Omega_S}} = \tilde \omega  \sqrt{\frac{d A_O}{d \Omega_S}} = \tilde \omega D.
\end{equation}
Since we are considering a light beam with a vertex at the source position, the initial conditions of the Sachs focusing equation are
\begin{equation}
D \left( \lambda_S \right) = 0, \qquad \left. \frac{d D\left( \lambda\right)}{d\lambda} \right|_{\lambda  = \lambda_S } =1,
\end{equation}
if we normalize $\lambda$ such that $\omega_S=1$. The  general initial conditions for an arbitrary affine 
parameter $\lambda$, are given by~(\ref{incond}). Choosing $\omega_S=1+z$ one obtains the luminosity distance
while $\omega_S=(1+z)^{-1}$ gives the angular diameter distance.

\section{Details for tensor perturbations}\label{app:ten}
Here we write down the nonvanishing Christoffel symbols and the Ricci tensor for the metric~(\ref{metricten}),
\begin{eqnarray}
\Gamma^0_{ij} &=& \eta^{0l} \left( H_{il,j} + H_{lj,i}- H_{ij,l} \right)  = \dot H_{ij} ,  \\
&&  \Rightarrow \quad \sum_i \Gamma^0_{ii} =0, \nonumber \\
\Gamma^i_{j0} &=& \eta^{ik} \left( H_{k0,j} + H_{jk,0} - H_{j0,k} \right) = \dot H_{ji}, \\
\nonumber && \Rightarrow \Gamma^i_{i0} =0, \\
\Gamma^i_{jl}&=& \eta^{ik} \left( H_{jk,l} + H_{kl,j} -H_{jl,k} \right)\nonumber \\
                &=&  H_{ji,l} + H_{il,j} -H_{jl,i}, \\
\Gamma^i_{ij} &=& H_{ij,i} + H_{ii,j} -H_{ij,i} = H_{ii,j}, \\
&& \Rightarrow\quad \Gamma^i_{ij} = 0. \nonumber
\end{eqnarray}
\begin{eqnarray}
R_{00} &=&R_{i0} \; = \; 0,\\
R_{ij} &=& \ddot H_{ij} -H_{ij,ll} = - \Box H_{ij}.
\end{eqnarray}
These components lead to
\begin{equation}
\mathcal R \!=\! \frac{1}{2} \left( R_{00} + 2 R_{i0} n^i + R_{ij} n^i n^j \right) \!=\! - \frac{1}{2} n^i n^j \Box H_{ij}.
\end{equation}

\section{Details for vector perturbations}
\subsection{Christoffel symbols and Ricci tensor}
\label{app:vec}
Here we write down the nonvanishing Christoffel symbols and the Ricci tensor for the metric~(\ref{metricvec}),
\begin{eqnarray}
\Gamma^i_{00}&=& - \dot B_{i}, \\
\Gamma^i_{j0} &=&  \frac{1}{2} \left( \partial_\eta \left( H_{i,j} +H_{j,i} \right) - B_{i,j} + B_{j,i} \right), \\
\Gamma^i _{i0} &=& \dot H_{i,i} \qquad \Rightarrow \quad  \Gamma^i_{i0}=0, \\
\Gamma^0_{ij} &=& \frac{1}{2} \left( B_{j,i} + B_{i,j} + \dot H_{i,j} + \dot H_{j,i}  \right) = \sigma_{ij}, \\
\Gamma^0_{ii} &=& B_{i,i} + \dot H_{i,i} \qquad \Rightarrow \quad \sum_i \Gamma^0_{ii} =0, \\
\Gamma^i_{ij} &= & H_{i,ij} \qquad \Rightarrow \quad  \Gamma^i_{ij} =0, \\
\Gamma^i_{jk} &=& H_{i,jk},\\
\Gamma^i_{ik} &=& H_{i,ik} \qquad \Rightarrow \quad  \Gamma^i_{ik} = 0,
\end{eqnarray}
\begin{eqnarray}
R_{00}& = &0, \\
R_{i0}&=&  \frac{1}{2} \left( B_{i,jj} + \dot H_{i,jj} \right) = \frac{1}{2} \nabla^2 \sigma_i, \\
R_{ij} &=& \frac{1}{2} \left( \dot B_{j,i} + \dot B_{i,j} + \ddot H_{i,j} + \ddot H_{j,i} \right) = \dot \sigma_{ij} .
\end{eqnarray}
With these Ricci tensor components we can easily compute
\begin{eqnarray}
\mathcal R &=& \frac{1}{2} \left( R_{00} + 2 R_{i0} n^i + R_{ij} n^i n^j \right) \nonumber \\
&=& \frac{1}{2} \left( \nabla^2 \sigma_i n^i + \dot \sigma_{ij} n^i n^j \right).
\end{eqnarray}

\subsection{The $c_\ell$ coefficients} \label{app:veccl}
We first derive in detail Eq.~(\ref{clvectorgen}). We use the relation 
\begin{equation} \label{e:vecadd}
Y_{1,\pm1}({ \bf n , \hat k} ) \!=\! \sqrt{\frac{4 \pi}{3}}\!\! \sum_{m' =-1}^1 \!\!Y_{1 m'} ({\bf  n,  e}){\;}^{\,}_{\mp 1}Y^*_{1m'}({\bf \hat k, e})\,.
\end{equation}
Here ${\,}^{\,}_{\mp 1}Y^*_{1m'}( {\bf \hat k,  e})$ is the vector spherical harmonic. The 
general addition theorem for spin weighted spherical harmonics used in Eq.~(\ref{e:vecadd}) above can be found, e.g., 
in~\cite{RD}. Applying this to Eq.~(\ref{46}), we obtain
\begin{eqnarray} 
n^i n^j  \sigma_{ij} &=& \frac{\left( 4 \pi \right)^2}{3} \sum_{\ell m} i^{\ell} \sum_{m' =-1}^{1} Y_{1m'} ({ \bf n}) Y_{\ell m}({ \bf n})  \nonumber\\
&& \hspace{-1cm} \times  \int \frac{d^3k}{\left( 2 \pi \right)^3}k j'_{\ell} ( k \Delta\eta)Y^*_{\ell m}({\bf \hat k}) \nonumber \\
&& \hspace{-0.4cm}  \left(\hat\sigma^+ {\,}^{\,}_{-1}Y^*_{1m'}({\bf \hat k}) - \hat\sigma^-{\,}^{\,}_{1} Y_{1m'}^*({\bf \hat k}) \right) .
\end{eqnarray}
Here we omit the arbitrary unit vector $\bf e$ in the notation, and we have introduced the helicity basis,
$$ \hat \si_i = \hat \si^+{ e}^{(+)}_i + \hat  \si^-{ e}^{(-)}_i $$
defined in Eq.~(\ref{e:hel}).  In the special case with 
${\bf n } = {\bf e}$ we obtain
\begin{eqnarray}
e^i e^j  \sigma_{ij} &=&  \frac{4 \pi}{ \sqrt{3} }\sum_{\ell} i^{\ell} \sqrt{2 \ell + 1} \nonumber\\
&&  \times  \int \frac{d^3 k }{\left( 2 \pi \right)^3} k j'_\ell \left( k \Delta\eta \right) Y^*_{\ell 0}({ \bf \hat k}) \nonumber \\
&& \qquad   \left(  \hat \sigma^+{\,}^{\,}_{-1}Y^*_{10} ({\bf \hat k}) - \hat \sigma^- {\,}^{\,}_{1}Y^*_{10}({\bf \hat k})  \right)  \!, \quad 
\end{eqnarray}
where we have used
\begin{equation} 
Y_{\ell m} \left( \bf e, \bf e \right) = \sqrt{\frac{2\ell +1}{4 \pi }} \delta_{m0},
\end{equation}
and Eq.~(\ref{e:vecadd}) for $\bf n = \bf e$ which yields,
\begin{eqnarray}
Y_{1, \pm 1 } ({\bf e , \hat k}) &=& \sqrt{\frac{4 \pi }{3}} \sum_{m'=-1}^1 
Y_{1m'}( {\bf e, \bf e}){\ }^{\,}_{\mp1}Y^*_{1m'} ({ \bf \hat k , \bf e}) \nonumber \\  
&=&{\,}^{\,}_{\mp 1} Y_{10}^*({ \bf \hat k , e}).
\end{eqnarray}
Since the two point correlation function~(\ref{48}) depends on the angle $\bf n \cdot n'$ only we can set $\bf n' = e $ without loss of generality. With this we find
\begin{eqnarray} \label{90}
&&\hspace{-0.8cm}\langle n^i n^j \sigma_{ij} e^i e^j \sigma_{ij}\rangle  \nonumber\\ \nonumber
&=&\frac{8}{3\sqrt{3}} \sum_{\ell,m} \sum_{\tilde \ell } i^{\ell - \tilde \ell} \sqrt{2 \tilde\ell + 1} \  Y_{\ell  m} \left( \bf n \right) \sum_{m'=-1}^1 Y_{1 m'} \left( \bf n \right)   \\ 
&&  \times \int dk \ k^4 j'_\ell( k \Delta\eta) j'_{\tilde\ell }( k \Delta\eta') P_\sigma \left( k \right) T_k ( \eta ) T_k ( \eta' ) \nonumber \\
&& \times \int d\Omega_{\bf \hat k} \ f_{m'} ({\bf  \hat k} )Y^*_{\ell m} ({\bf \hat k} ) Y_{\tilde\ell 0} ({\bf \hat k} ), 
 \end{eqnarray}
 where we have introduced
 \begin{equation}
f_{m'}( {\bf  \hat k}) \!=\! \! {~}^{~} _{-1} Y_{1 m' }^*({\bf  \hat k}) \ _{-1} Y_{10} ({\bf  \hat k} ) \!+\!  
 {~}^{~}_1 Y_{1 m'}^*({\bf\hat k }) \ _{1} Y_{10}({\bf\hat k} ) .
\end{equation}
Since the spherical harmonics form an orthogonal basis on $S_2$, we can expand the product of two of them again in terms of spherical harmonics using the Clebsch--Gordan coefficients~\cite{CG}. In the case of Eq.~(\ref{90}) we use
\begin{eqnarray}
&&\hspace{-1.2cm}Y_{\ell m}^*({\bf  \hat k}) Y_{\tilde \ell 0 }({\bf\hat k }) = \sum_{L=|\ell-\ell'|}^{\ell+\ell'} \sqrt{\frac{\left( 2 \ell +1  \right) \left( 2 \tilde\ell +1 \right) }{4 \pi \left( 2L +1 \right) }}  \nonumber   \\ \label{C20}
&&~ \times \langle\ell ,0 , \tilde\ell , 0 | L, 0\rangle\langle\ell , m , \tilde\ell , 0 | L, m \rangle Y_{Lm}^*({\bf\hat k}).
\end{eqnarray}
The dependence of the spherical harmonics on the azimuthal angle $\phi$, 
\begin{equation}
Y^*_{Lm} \propto e^{-im\phi} \qquad \text{and} \qquad {\,}^{\,}_{\pm 1} Y^*_{1m '} \propto e^{-im ' \phi}
\end{equation}
implies that the integral over angles in Eq.~(\ref{90}) only contributes for $m' = -m $. Therefore also $m \in \left\{ -1,0, 1 \right\}$. Since $f_{m'}( \bf \hat k)$ contains only terms that either do not depend on $\theta$ or that are 
quadratic in $\sin \left( \theta \right)$ and $\cos( \theta)$ the only nonvanishing contributions are $L=0$ or $L=2$. Analogous to Eq.~(\ref{C20}) we write
\begin{eqnarray}
&&\hspace{-1.0cm}Y_{\ell,m} \left( \bf n \right) Y_{1 ,-m} \left( \bf n \right) = \sum_{n} \sqrt{\frac{3 \left( 2\ell + 1\right)}{4 \pi( 2n +1)}}\nonumber \\ 
&&  \times \langle\ell ,0,1,0|n,0\rangle\langle \ell, m, 1 ,-m | n, 0\rangle Y_{n0} \left( \bf n \right)  .
\end{eqnarray}
The addition theorem for the spherical harmonics implies
\begin{equation}
P_{n } \left( \bf n \cdot e \right) = \sqrt{\frac{4 \pi}{2n +1 }} Y_{n 0 } \left( \bf n \right).
\end{equation}
Using these identities we can rewrite the correlation function~(\ref{90}) as
\begin{eqnarray}
&& \hspace{-0.3cm} \langle n^i n^j \sigma_{ij} e^i e^j \sigma_{ij}\rangle= \sum_{n} \sum_{L=0,2} \sum_{\ell , \tilde\ell } 
\sum_{m= -1}^{1} \frac{i^{\ell - \tilde \ell }}{3 \pi^{3/2}}  \nonumber\\ 
&& \quad \times \left( 2\ell +1 \right) \left( 2 \tilde\ell +1 \right) \left( 2 L +1 \right)^{-1/2} \nonumber \\
&& \quad \times  \langle\ell ,0 , \tilde\ell , 0 | L, 0\rangle\langle \ell , m , \tilde\ell , 0 | L, m \rangle  \nonumber \\
&& \quad \times  \langle\ell ,0,1,0| n,0 \rangle  \langle\ell, m, 1 ,-m | n, 0\rangle  \nonumber \\
&& \hspace{-0.5cm} \times \! \!\int \!\!dk  k^4 j'_\ell( k \Delta\eta) j'_{\tilde\ell } ( k \Delta\eta') P_\sigma ( k ) T_k (\eta)T_k(\eta')B_{Lm}P_n \! \left( \bf n \! \cdot \! e \right),\nonumber \\ &&
\end{eqnarray}
where we have introduced
\begin{equation}
B_{Lm} = \int d \Omega_{\bf \hat k} \  f_{-m}( {\bf \hat k}) Y_{Lm }^*( {\bf \hat k}). 
\end{equation}
The nonvanishing coefficients are given by
\begin{equation}
B_{00}= \frac{1}{\sqrt{\pi}}, \quad B_{2,\pm1} =  \frac{1}{2} \sqrt{\frac{3}{5 \pi}}, \quad B_{2,0} = - \frac{1}{\sqrt{5\pi}}.
\end{equation}
Computing the sum of the Clebsch--Gordan coefficients, we find
\begin{eqnarray}
&&\hspace{-1.0cm}\langle n^i n^j \sigma_{ij} e^i e^j \sigma_{ij}\rangle= \sum_{\ell} \frac{\ell(\ell+1)}{2\pi(2\ell+1)} 
         P_\ell\left( \bf n \cdot e \right) \nonumber \\
 && \hspace{-0.8cm} \times \int dk \ k^4 P_\sigma( k ) T_k( \eta) T_k( \eta' ) \nonumber   \\
 &&  \hspace{-0.8cm} \times \left[ j'_{\ell \!-\! 1}( k\Delta\eta) j'_{\ell \!-\! 1} ( k\Delta\eta' ) + j'_{\ell \!+ \!1}( k\Delta\eta)j'_{\ell\!+\!1}( k\Delta\eta') \right] \!.
\end{eqnarray}
Using this result for the $\bar{c}_\ell$'s defined in Eq.~(\ref{48}), we obtain directly Eq.~(\ref{clvectorgen}).

\onecolumngrid
\vspace{1cm}
Now we compute  the full $c_\ell$ coefficient defined in~(\ref{29b}). We use the Limber approximation 
(Appendix~\ref{app:lim}) to do the time integrals $\int d\eta T_k(\eta) j'_{\ell - 1} ( k(\eta_O-\eta) )$. We start with the Doppler terms, the first line of Eq.~(\ref{distvec2}), which contribute to the dipole term only
\begin{eqnarray}
c^{\text{D}}_1&=& \frac{4 \pi}{3} \frac{\mathcal{H}_S^{-1}}{\Delta\eta_S}  \frac{\mathcal{H}_{S'}^{-1}}{\Delta\eta_{S'}} <\left| \vec v_O \right|^2 >.
 \end{eqnarray}
This dipole is the same as the one from the scalar analysis~\cite{BDG}. We cannot, of course, decide which part of the observer velocity comes from scalar perturbations and which part from vector perturbations. Since this dipole term is highly nonlinear, we neglect it in the subsequent analysis and consider only $\ell\ge 2$.
We now determine the other terms. From the peculiar velocity of the source we obtain
\bea
c_\ell^{(1)} = \frac{2 \ell \left( \ell +1 \right)}{\pi \left( 2 \ell +1 \right)^2} \left( 1 - \frac{\mathcal H_S^{-1}}{\Delta\eta_S} \right)\left( 1 - \frac{\mathcal H_{S'}^{-1}}{\Delta\eta_{S'}} \right) \int dk \ k^2 P_v \left( k \right) T^v_k \left( \eta_S \right) T^v_k \left( \eta_{S'} \right)\! \!  \sum_{\tilde \ell= \ell-1, \ell +1} \! \! \! j_{\tilde \ell} \left( k \Delta \eta_S \right) j_{\tilde \ell} \left( k \Delta \eta_{S'} \right) ,
\eea
where we have introduced the velocity power spectrum defined by
\be
\langle \hat v^{\pm}\left( \eta_S, \mathbf k\right) \hat v^{\pm*} \left( \eta_{S'}, \mathbf k'\right) \rangle= \left( 2 \pi \right)^3  \delta^{(3)} \left( \mathbf k -\mathbf {k'} \right) P_v \left( k \right) T_k^v \left( \eta_S \right) T_k^v\left( \eta_{S'} \right) 
\ee
and $\hat v^{\pm}$ is the peculiar velocity in terms of the helicity basis defined in Eq.~(\ref{e:hel}).

The second line of the redshift--distance relation~(\ref{distvec2}) leads to
\begin{eqnarray}
c_\ell^{(2)} &\cong& \frac{\ell \left( 1 + \ell \right)}{2 \pi \left( 1 + 2 \ell \right)^2 }\frac{\mathcal{H}_S^{-1}}{\Delta \eta_S} \frac{\mathcal{H}_{S'}^{-1}}{\Delta \eta_{S'}} \int dk \ k^2 P_\sigma \left( k \right) \nonumber  \\ 
&& \hspace{-0.5cm}\! \! \times  \! \! \! \!    \sum_{\tilde \ell = \ell \! -\! 1, \ell \!+ \! 1} \! \! \left[ I_{ \tilde \ell -1 }^2 T_k^2 \!  \left( \eta_{\tilde \ell -1,k} \right) \Theta \left( \eta_{ \tilde \ell -1,k} \!- \! \eta_S \right) \Theta \left( \eta_{\tilde \ell -1,k} \! -\!  \eta_{S'} \right)  + \left( \frac{ \tilde \ell +1 }{\tilde \ell }\right)^2I_{\tilde \ell}^2 T^2_k \! \left( \eta_{\tilde \ell,k} \right) \Theta \left( \eta_{\tilde \ell,k} \! -\!  \eta_S \right) \Theta \left(\eta_{\tilde \ell,k} \! - \! \eta_{S'} \right)     \right. \nonumber \\
&& \qquad \hspace{-0.6cm}  - \left. \frac{ \tilde \ell +1 }{ \tilde \ell }  I_{ \tilde \ell -1 }  I_{\tilde \ell} T_k \! \left(  \eta_{\tilde \ell -1,k}  \right) T_k  \! \left(  \eta_{\tilde \ell ,k}  \right)  \left( \Theta \left( \eta_{ \tilde \ell ,k} \!- \! \eta_S \right) \Theta \left( \eta_{ \tilde \ell -1,k} \!- \! \eta_{S'} \right) + \Theta \left( \eta_{ \tilde \ell -1,k} \!- \! \eta_S \right)\Theta \left( \eta_{ \tilde \ell ,k} \!- \! \eta_{S'} \right)\right) \right] \! .
\end{eqnarray}
Analogously the third line yields
\begin{eqnarray}
c_\ell^{(3)} &\cong& \frac{\ell \left( 1 + \ell \right)}{2 \pi \left( 1 + 2 \ell \right)^2 }\frac{1}{\Delta\eta_S} \frac{1}{\Delta\eta_{S'}} \int dk \ P_\sigma \left( k \right) \nonumber \\
&& \hspace{-0.5cm}\! \!  \times  \! \! \!    \sum_{\tilde \ell = \ell \! -\! 1, \ell \!+ \! 1} \!\!\! \left[ \left( \tilde \ell \! -1 \! \right)^2 \! I_{\tilde \ell \!- \! 1}^2   T_k^2  \! \left(  \eta_{\tilde \ell -1 ,k}  \right) \Theta \!  \left( \eta_{ \tilde \ell \! -\! 1 ,k} \!- \! \eta_S \right) \Theta \!  \left( \eta_{ \tilde \ell \! -\! 1 ,k} \!- \! \eta_{S'} \right)  \!+\! \left( \tilde \ell \! +\! 1 \right)^2 \! I_{\tilde \ell}^2    T_k^2  \! \left(  \eta_{\tilde \ell ,k}  \right) \Theta \!   \left( \eta_{ \tilde \ell ,k} \!- \! \eta_S \right) \Theta \!  \left( \eta_{ \tilde \ell ,k} \!- \! \eta_{S'} \right)    \right. \nonumber \\
&& \qquad \hspace{-0.6cm}  \left. \!  - \!  \left( \tilde \ell^2 \!- \! 1 \right) I_{ \tilde \ell \! -\! 1 } I_{ \tilde \ell } T_k  \! \left(  \eta_{\tilde \ell -1 ,k}  \right) T_k  \! \left(  \eta_{\tilde \ell ,k}  \right) \left( \Theta \!   \left( \eta_{ \tilde \ell -1 ,k} \!- \! \eta_S \right)   \Theta \!   \left( \eta_{ \tilde \ell ,k} \!- \! \eta_{S'} \right) +  \Theta \!   \left( \eta_{ \tilde \ell ,k} \!- \! \eta_S \right)  \Theta \!   \left( \eta_{ \tilde \ell -1 ,k} \!- \! \eta_{S'} \right)  \right) \right] \! . 
\end{eqnarray}
The fourth line is composed of the two terms that contribute to ${\mathcal R}$ given in Eq.~(\ref{e:Rvec}). Denoting them with superscripts $(41)$, $(42)$, and their correlation with   $(412)$ we obtain
\begin{eqnarray}
c^{(41)}_\ell  \!\!&\cong& \!\! \frac{\ell \left( 1 \!+\! \ell \right)}{8 \pi \left( 1 \!+ \!2 \ell \right)^2 }\!\frac{1}{\Delta \eta_S}\! \frac{1}{\Delta\eta_{S'}} \! \!\!\int \!\! dk \ k^2 P_\sigma\! \left( k \right)  \!\!\!\!\!\!  \sum_{\tilde \ell = \ell \! -\! 1, \ell \!+ \! 1} \!\!\! \! \left[ \tilde \ell^2 \! \left( \eta_{\tilde \ell,k} \! -\! \eta_S\right) \! \left( \eta_{\tilde \ell,k} \! - \! \eta_{S'}  \right) \!  I_{\tilde \ell}^2  T_k^2 \! \left( \eta_{\tilde \ell,k} \right) \! \Theta \! \left( \eta_{\tilde \ell,k} \!- \! \eta_S \right)\! \Theta \! \left(  \eta_{\tilde \ell,k}\! -\!\eta_{S'} \right) \right]   \! \!,  \hspace{0.9cm} \\
c^{(42)}_\ell \!\!&\cong&\frac{\ell \left( 1 + \ell \right)}{8 \pi \left( 1 + 2 \ell \right)^2 } \frac{1}{\Delta \eta_S} \frac{1}{\Delta \eta_{S'}} \int dk \ P_\sigma \left( k \right)\nonumber \\
&& \hspace{-0.5cm} \! \!  \times  \! \! \! \!    \sum_{\tilde \ell = \ell \! -\! 1, \ell \!+ \! 1} \!\!\!   \left[ \left( \eta_{\tilde \ell -1,k} - \eta_S  \right) \left( \eta_{\tilde \ell -1 , k} - \eta_{S'}  \right) \left( \tilde \ell  -1 \right)^2 I_{\tilde \ell -1}^2\dot T_k^2 \left( \eta_{\tilde \ell -1,k} \right) \Theta \left(\eta_{\tilde \ell -1,k} - \eta_S  \right)  \Theta \left( \eta_{\tilde \ell -1,k} - \eta_{S'} \right)   \right. \nonumber \\
&&    + \left( \eta_{\tilde \ell ,k} - \eta_S  \right) \left( \eta_{\tilde \ell  , k} - \eta_{S'}  \right) \left( \tilde \ell  + 1 \right)^2   I_{\tilde \ell }^2\dot T_k^2 \left( \eta_{\tilde \ell -1,k} \right) \Theta \left(\eta_{\tilde \ell ,k} - \eta_S  \right)  \Theta \left( \eta_{\tilde \ell ,k} - \eta_{S'} \right)       \nonumber \\
&& -  \left( \tilde \ell^2 - 1 \right) I_{\tilde \ell -1}  I_{\tilde \ell } \dot T_k \left( \eta_{\tilde \ell - 1,k} \right) \dot T_k \left(\eta_{ \tilde \ell,k} \right) \left( \left( \eta_{\tilde \ell -1,k} - \eta_S  \right) \left( \eta_{\tilde \ell ,k} - \eta_{S'}  \right) \Theta\left( \eta_{\tilde \ell -1,k} - \eta_S  \right) \Theta \left( \eta_{\tilde \ell ,k} - \eta_{S'}  \right) \right. \nonumber \\
&& \qquad \qquad \qquad \qquad \qquad \qquad \qquad \qquad   \left. \left. +  \left( \eta_{\tilde \ell,k} - \eta_S  \right) \left( \eta_{\tilde \ell -1,k} - \eta_{S'}  \right) \Theta \left( \eta_{\tilde \ell,k} - \eta_S  \right) \Theta \left( \eta_{\tilde \ell -1,k} - \eta_{S'}  \right) \right) \right],  \\
c^{(412)}_\ell &\cong&  \frac{\ell \left( 1 + \ell \right)}{8 \pi \left( 1 + 2 \ell \right)^2 }\frac{1}{\Delta \eta_S} \frac{1}{\Delta \eta_{S'}}\int dk \ k \ P_\sigma \left( k \right) \nonumber  \\
&& \! \! \! \! \times  \! \! \! \!    \sum_{\tilde \ell = \ell \! -\! 1, \ell \!+ \! 1} \!\!\! \!\left[ \left( \eta_{\tilde \ell,k} \! - \! \eta_S \right) \left( \eta_{\tilde \ell - 1,k} \! -\! \eta_{S'}  \right) \tilde \ell \left( \tilde \ell -1 \right) I_{ \tilde \ell -1  }  I_{\tilde \ell} \dot T_k \left( \eta_{\tilde \ell -1,k} \right) T_k \left( \eta_{\tilde \ell,k} \right) \Theta \left( \eta_{\tilde \ell,k} \! - \! \eta_S \right) \Theta \left( \eta_{\tilde \ell -1 ,k} \! -\! \eta_{S'} \right) \right.  \nonumber\\
&& \qquad \quad  \left. - \left( \eta_{\tilde \ell,k} \! - \! \eta_S  \right) \left( \eta_{\tilde \ell,k}\! -\! \eta_{S'} \right)\tilde \ell \left( \tilde \ell+1 \right) I_{\tilde \ell}^2 T_k \left( \eta_{ \tilde \ell,k} \right) \dot T_k \left( \eta_{ \tilde \ell,k} \right)  \Theta \left( \eta_{\tilde \ell, k}\! - \!\eta_S \right) \Theta \left( \eta_{\tilde \ell,k}\! - \!\eta_{S'}  \right) \right]  \nonumber \\ &&  + \left( S\leftrightarrow S' \right)  .
\end{eqnarray}
Next, we compute the cross terms between the different lines of the distance--redshift relation~(\ref{57b}). We start with
 the second and third lines,
\begin{eqnarray}
c^{(23)}_\ell &\cong& - \frac{\ell \left( 1 + \ell \right)}{2 \pi \left( 1 + 2 \ell \right)^2 }\frac{\mathcal{H}_S^{-1}}{\Delta \eta_S} \frac{1}{\Delta \eta_{S'}} \int dk \ k \ P_\sigma \left( k \right) \nonumber  \\
&& \! \! \! \! \times  \! \! \! \! \!   \sum_{\tilde \ell = \ell \! -\! 1, \ell \!+ \! 1} \!\!  \left[ \!\left( \tilde \ell \! -\! 1 \right) I_{\tilde \ell \!-\!1}^2 T_k^2 \! \left( \eta_{ \tilde \ell \!-\!1 ,k} \right) \Theta \! \left( \eta_{\tilde \ell \!-\!1,k} \!-\!\eta_S \right) \Theta \!\left( \eta_{ \tilde \ell \!-\!1,k} \!-\! \eta_{S'}  \right)   + \frac{\left( \tilde \ell \!+\!1 \right)^2}{\tilde \ell } I_{\tilde \ell }^2 T_k^2 \!\left( \eta_{\tilde \ell,k} \right)  \Theta\! \left(  \eta_{\tilde \ell,k}\! -\! \eta_S \right) \Theta \left( \eta_{ \tilde \ell,k} \!-\! \eta_{S'} \right)  \right. \nonumber \\
&& \qquad \qquad   - \left( \tilde \ell +1 \right)I_{\tilde \ell -1 } I_{ \tilde \ell } T_k \left( \eta_{\tilde \ell -1,k} \right)  T_k \left( \eta_{\tilde \ell,k} \right) \Theta \left(  \eta_{\tilde \ell -1 ,k}-\eta_S \right) \Theta \left( \eta_{ \tilde \ell,k} -\eta_{S'} \right) \nonumber \\
&& \qquad \qquad \qquad \left. - \frac{\tilde \ell^2 -1 }{\tilde \ell } I_{\tilde \ell -1 } I_{\tilde \ell } T_k \left( \eta_{\tilde \ell -1,k} \right) T_k \left( \eta_{ \tilde \ell,k} \right) \Theta \left( \eta_{ \tilde \ell,k} - \eta_S \right) \Theta  \left(  \eta_{\tilde \ell -1,k} -\eta_{S'} \right) \right] \qquad + \left( S\leftrightarrow S' \right) . 
\end{eqnarray}
The second and fourth lines yield
\begin{eqnarray}
c^{(241)}_\ell &\cong& \frac{\ell \left( 1 + \ell \right)}{4 \pi \left( 1 + 2 \ell \right)^2 }\frac{\mathcal{H}_S^{-1}}{\Delta \eta_S} \frac{1}{\Delta \eta_{S'}}\int dk \ k^2 P_\sigma \left( k \right) \nonumber   \\
&& \! \! \! \! \times  \! \! \! \!    \sum_{\tilde \ell = \ell \! -\! 1, \ell \!+ \! 1} \!\!\!   \left[ \tilde \ell \left( \eta_{\ell,k} - \eta_{S'} \right) I_{\tilde \ell -1 }  I_{\tilde \ell} T_k \left( \eta_{ \tilde \ell-1,k} \right) T_k \left( \eta_{\tilde \ell,k} \right) \Theta \left( \eta_{ \tilde \ell -1,k} -  \eta_S \right) \Theta \left( \eta_{\tilde \ell,k} -  \eta_{S'}  \right) \nonumber \right. \\
&&\left.  \qquad  - \left( \tilde \ell +1 \right) \left( \eta_{\tilde \ell,k} - \eta_{S'}  \right) I_{\tilde \ell }^2 T^2_k \left( \eta_{ \tilde \ell,k} \right) \Theta \left( \eta_{\tilde \ell,k} - \eta_S  \right) \Theta \left( \eta_{ \tilde \ell,k} - \eta_{S'} \right) \right]   \qquad + \left( S\leftrightarrow S' \right),  \\  
c^{(242)}_\ell  &\cong& \frac{\ell \left( 1 + \ell \right)}{4 \pi \left( 1 + 2 \ell \right)^2 }\frac{\mathcal{H}_S^{-1}}{\Delta \eta_S} \frac{1}{\Delta \eta_{S'}} \int dk \ k P_\sigma \left( k \right) \nonumber  \\
&& \! \! \! \! \times  \! \! \! \!    \sum_{\tilde \ell = \ell \! -\! 1, \ell \!+ \! 1} \!\!\!    \left[ \left( \eta_{\tilde \ell-1,k} - \eta_{S'}\right) \left( \tilde \ell -1 \right)  I_{\tilde \ell -1 }^2 T_k \left( \eta_{ \tilde \ell -1,k} \right)   \dot  T_k \left( \eta_{ \tilde \ell -1} \right)  \Theta \left( \eta_{\tilde \ell -1,k} - \eta_S \right) \Theta \left(\eta_{ \tilde \ell -1,k} -\eta_{S'}  \right) \nonumber \right. \\
 && \qquad \quad - \left( \eta_{\tilde \ell,k} - \eta_{S'}  \right) \left( \tilde \ell +1 \right)  I_{\tilde \ell -1 } I_{ \tilde \ell } T_k \left( \eta_{ \tilde \ell -1,k} \right)  \dot T_k \left( \eta_{\tilde \ell,k} \right) \Theta \left( \eta_{\tilde \ell -1,k} - \eta_S  \right) \Theta \left( \eta_{\tilde \ell,k} - \eta_{S'}  \right) \nonumber \\
  && \qquad \quad  - \left( \eta_{\tilde \ell -1 ,k} - \eta_{S'}\right) \frac{ \tilde \ell^2 -1}{\tilde \ell } I_{\tilde \ell -1 }  I_{\tilde \ell} \dot T_k \left( \eta_{ \tilde \ell -1,k} \right) T_k \left( \eta_{\tilde \ell,k} \right)  \Theta \left( \eta_{\tilde \ell,k} - \eta_S  \right) \Theta \left( \eta_{\tilde \ell -1 ,k} -\eta_{S'} \right) \nonumber \\
   &&   \qquad \quad + \left. \left( \eta_{\tilde \ell,k} - \eta_{S'} \right) \frac{\left( \tilde \ell +1 \right)^2}{\tilde \ell} I_{\tilde \ell  }^2 T_k \left( \eta_{ \tilde \ell,k}  \right)  \dot T_k \left( \eta_{\tilde \ell,k}  \right) \Theta \left( \eta_{ \tilde \ell,k} - \eta_S  \right) \Theta \left( \eta_{\tilde \ell,k} - \eta_{S'}   \right)  \right]   + \left( S\leftrightarrow S' \right) ,\quad
\end{eqnarray}
and, the third and fourth lines give
\begin{eqnarray}
c^{(341)}_\ell&\cong& -  \frac{\ell \left( 1 + \ell \right)}{4 \pi \left( 1 + 2 \ell \right)^2 }\frac{1}{\Delta \eta_S} \frac{1}{\Delta \eta_{S'}} \int dk \ k \ P_\sigma \left( k \right) \nonumber  \\
&& \! \! \! \! \times  \! \! \! \!    \sum_{\tilde \ell = \ell \! -\! 1, \ell \!+ \! 1} \!\!\!  \left[ \left( \eta_{\tilde \ell,k} - \eta_{S'}  \right) \tilde \ell \left( \tilde \ell-1 \right)  I_{\tilde \ell -1 }  I_{ \tilde \ell } T_k \left( \eta_{\tilde \ell -1,k} \right)  T_k \left( \eta_{\tilde \ell,k} \right) \Theta \left( \eta_{\tilde \ell -1,k} - \eta_S \right) \Theta \left( \eta_{\tilde \ell,k} - \eta_{S'}  \right) \right. \nonumber \\
  && \qquad \quad - \left. \left( \eta_{\tilde \ell,k} - \eta_{S'} \right) \tilde \ell \left( \tilde \ell +1 \right) I_{ \tilde \ell }^2 T_k^2 \left( \eta_{ \tilde \ell,k} \right) \Theta \left( \eta_{\tilde \ell,k} - \eta_S  \right)  \Theta \left( \eta_{\tilde \ell ,k} -\eta_{S'} \right) \right]  \qquad + \left( S\leftrightarrow S' \right) , \\
c^{(342)}_\ell&\cong&- \frac{\ell \left( 1 + \ell \right)}{4 \pi \left( 1 + 2 \ell \right)^2 }\frac{1}{\Delta \eta_S} \frac{1}{\Delta \eta_{S'}} \int dk \  P_\sigma \left( k \right) \nonumber  \\
&& \! \! \! \! \times  \! \! \! \!    \sum_{\tilde \ell = \ell \! -\! 1, \ell \!+ \! 1} \!\!\! \left[  \left( \tilde \ell -1 \right)^2 \left( \eta_{\tilde \ell-1,k} -\eta_{S'} \right)I_{\tilde \ell -1}^2 T_k \left( \eta_{ \tilde \ell -1,k} \right) \dot T_k \left( \eta_{\tilde \ell -1,k} \right) \Theta \left(\eta_{\tilde \ell -1,k}- \eta_S  \right)  \Theta \left( \eta_{\tilde \ell -1,k} - \eta_{S'}  \right) \right. \nonumber \\
&& \qquad \quad - \left( \tilde \ell^2  -1 \right) \left( \eta_{\tilde \ell,k}-\eta_{S'}  \right) I_{\tilde \ell -1 }  I_{\tilde \ell } T_k \left( \eta_{\tilde \ell -1,k} \right) \dot T_k \left( \eta_{\tilde \ell,k} \right) \Theta \left(\eta_{\tilde \ell -1,k} - \eta_S  \right) \Theta \left( \eta_{\tilde \ell,k}- \eta_{S'}  \right)  \nonumber \\
&& \qquad  \quad - \left( \tilde \ell^2 -1 \right) \left( \eta_{\tilde \ell -1 ,k} - \eta_{S'}  \right) I_{\tilde \ell -1 }  I_{ \tilde \ell } \dot T_k \left( \eta_{ \tilde \ell -1,k} \right) T_k \left( \eta_{ \tilde \ell,k} \right) \Theta \left( \eta_{\tilde \ell,k}-\eta_S  \right) \Theta \left( \eta_{\tilde \ell -1 ,k} - \eta_{S'} \right])   \nonumber \\
&& \qquad  \quad + \left.  \left(  \tilde \ell +1 \right)^2 \left( \eta_{\tilde \ell,k} - \eta_{S'}  \right) I_{\tilde \ell }^2 T_k \left( \eta_{\tilde \ell,k} \right) \dot T_k \left( \eta_{\tilde \ell,k} \right) \Theta \left( \eta_{\tilde \ell,k} - \eta_S \right) \Theta \left( \eta_{\tilde \ell,k} - \eta_{S'}  \right) \right]  + \left( S\leftrightarrow S' \right) . \quad 
\end{eqnarray}

To determine the correlation between the peculiar velocities and the shear on the constant time hypersurface $\sigma_{ij}$ we need to specify the gravitation theory. We choose GR by using the Einstein's equations~(\ref{vecEinstein}). Correlating the term for the peculiar velocity of the source $v_S$ with the others, we find 
\begin{eqnarray}
c_\ell^{(12)} \!\! &= & \!\! \frac{-1}{16 \pi G a_S^2 \left( \bar \rho_S + \bar p_s \right)}\! \left( 1\! -\! \frac{\mathcal H_S^{-1}}{\Delta \eta_S}\right)\!\frac{\mathcal{H}_{S'}^{-1}}{\Delta\eta_{S'}}\! \frac{2 \ell \left( \ell+1 \right) }{\pi \left( 2 \ell +1 \right)^2}\!\!\int \!\! dk  k^5 P_\sigma\! \left( k \right) T_k\! \left( \eta_S \right) \!\! \!\!\sum_{\tilde \ell = \ell\!-\!1, \ell \!+\!1 }\!\! \!\! j_{\tilde \ell  } \left( k \Delta \eta_S \right)  \int_{\eta_{S'}}^{\eta_O} \!\! d\eta T_k \left( \eta \right)j'_{\tilde \ell} \left( k \Delta \eta \right) \nonumber   \\
&& \qquad +  \left( S\leftrightarrow S' \right) \nonumber \\
&\cong&   \!\! \frac{-1}{16 \pi G a_S^2 \left( \bar \rho_S + \bar p_s \right)}\! \left( 1\! -\! \frac{\mathcal H_S^{-1}}{\Delta \eta_S}\right)\!\frac{\mathcal{H}_{S'}^{-1}}{\Delta\eta_{S'}}\! \frac{2 \ell \left( \ell+1 \right) }{\pi \left( 2 \ell +1 \right)^2}  \!\! \!\!\sum_{\tilde \ell = \ell\!-\!1, \ell \!+\!1 } \left[ \frac{I_{\tilde \ell} \tilde \ell^4}{\Delta \eta_S^5}  P_\sigma \left( \frac{\tilde \ell}{\Delta \eta_S} \right) T_{\tilde \ell / \Delta \eta_S}\!\left( \eta_S \right)\right. \nonumber \\
&& \left. \times \left( T_{\tilde \ell / \Delta \eta_S} \! \left( \eta_S + \frac{\Delta \eta_S}{\tilde \ell}\right) I_{\tilde \ell -1 } \Theta \left( \eta_S - \eta_{S'} + \frac{\Delta \eta_S}{\tilde \ell}\right)  -  T_{\tilde \ell / \Delta \eta_S} \! \left( \eta_S\right) I_{\tilde \ell } \frac{\tilde \ell +1 }{\tilde \ell} \Theta \left( \eta_S - \eta_{S'} \right)  \right) \right] +  \left( S\leftrightarrow S' \right)  , \\
c_\ell^{(13)} \!\! &= &\! \! \frac{1}{16 \pi G a_S^2 \left( \bar \rho_S \!+ \!\bar p_s \right)} \!\left( 1\! -\! \frac{\mathcal H_S^{-1}}{\Delta \eta_S}\right)\!\frac{1}{\Delta\eta_{S'}} \!\frac{2 \ell \left( \ell\!+\!1 \right) }{\pi \left( 2 \ell\! +\!1 \right)^2}\! \!\!\int \! \!dk  k^5 P_\sigma \!\left( k \right) T_k\! \left( \eta_S \right) \!\!\!\! \sum_{\tilde \ell = \ell\!-\!1, \ell \!+\!1 } \!\!\!\! j_{\tilde \ell  } \left( k \Delta \eta_S \right) \!\! \int_{\eta_{S'}}^{\eta_O} \!\! d\eta \left( \eta_O \!- \!\eta \right) T_k\! \left( \eta \right)j'_{\tilde \ell} \left( k \Delta \eta \right)  \nonumber \\
&& \qquad +  \left( S\leftrightarrow S' \right) \nonumber \\
&\cong&   \! \! \frac{1}{16 \pi G a_S^2 \left( \bar \rho_S \!+ \!\bar p_s \right)} \!\left( 1\! -\! \frac{\mathcal H_S^{-1}}{\Delta \eta_S}\right)\!\frac{1}{\Delta\eta_{S'}} \!\frac{2 \ell \left( \ell\!+\!1 \right) }{\pi \left( 2 \ell\! +\!1 \right)^2}\!\! \sum_{\tilde \ell = \ell\!-\!1, \ell \!+\!1 } \!\! \left[ \frac{I_{\tilde \ell} \tilde \ell^3}{\Delta\eta_S^4} P_\sigma \left( \frac{\tilde \ell }{\Delta \eta_S} \right) T_{\tilde \ell / \Delta \eta_S}\! \left( \eta_S \right)           \right. \nonumber\\
&&\hspace{-0.5cm} \left. \times\! \left( T_{\tilde \ell / \Delta \eta_S} \! \left( \eta_S \!+\! \frac{\Delta \eta_S}{\tilde \ell}\!\right) I_{\tilde \ell \!-\!1 } \left(\! \tilde \ell \!-\!1\! \right) \Theta \!\left(\! \eta_S \!-\! \eta_{S'}\! +\! \frac{\Delta \eta_S}{\tilde \ell}\!\right) \! -\!  T_{\tilde \ell / \Delta \eta_S} \! \left( \eta_S\right) I_{\tilde \ell } \left( \tilde \ell \!+\!1 \right)\Theta \!\left( \eta_S \!-\! \eta_{S'} \right)  \right) \right] +  \left( S\leftrightarrow S' \right),   \\
c_\ell^{(141)} \!\! &= & \!\! \frac{1}{8 \pi G a_S^2 \left( \bar \rho_S \!+\! \bar p_s \right)} \!\left( 1\! -\! \frac{\mathcal H_S^{-1}}{\Delta \eta_S}\right)\!\frac{1}{\Delta\eta_{S'}} \frac{2 \ell \left( \ell+1 \right) }{\pi \left( 2 \ell +1 \right)^2} \nonumber \\ 
&& \qquad \times \int \! dk \ k^6 P_\sigma \left( k \right) T_k \left( \eta_S \right) \!\! \sum_{\tilde \ell = \ell\!-\!1, \ell \!+\!1 } \!\! j_{\tilde \ell  } \left( k \Delta \eta_S \right)  \int_{\eta_{S'}}^{\eta_O} \!\! d\eta\! \left( \eta_O \!- \!\eta \right)\! \left( \eta \!- \!\eta_{S'} \right)\! T_k \left( \eta \right)j'_{\tilde \ell} \left( k \Delta \eta \right)  \qquad +  \left( S\leftrightarrow S' \right) \nonumber \\
&\cong&   \!\! \frac{1}{8 \pi G a_S^2 \left( \bar \rho_S \!+\! \bar p_s \right)} \!\left( 1\! -\! \frac{\mathcal H_S^{-1}}{\Delta \eta_S}\right)\!\frac{1}{\Delta\eta_{S'}} \frac{2 \ell \left( \ell+1 \right) }{\pi \left( 2 \ell +1 \right)^2}  \!\! \sum_{\tilde \ell = \ell\!-\!1, \ell \!+\!1 } \!\!  \left[ \frac{I_{\tilde \ell} \tilde \ell^4 }{\Delta\eta_S^5} P_\sigma \left( \frac{\tilde \ell }{\Delta \eta_S}\right) T_{\tilde \ell /\Delta \eta_S}\! \left( \eta_S \right)    \nonumber \right. \\
&&\hspace{-0.5cm}  \times\! \left( T_{\tilde \ell / \Delta \eta_S} \! \left( \eta_S \!+\! \frac{\Delta \eta_S}{\tilde \ell}\!\right) I_{\tilde \ell \!-\!1 } \left(\! \tilde \ell \!-\!1\! \right)\left(\! \eta_S \!-\! \eta_{S'}\! +\! \frac{\Delta \eta_S}{\tilde \ell}\!\right) \Theta \!\left(\! \eta_S \!-\! \eta_{S'}\! +\! \frac{\Delta \eta_S}{\tilde \ell}\!\right) \nonumber \right. \\
&&\left.\left.  - T_{\tilde \ell / \Delta \eta_S} \! \left( \eta_S\right) I_{\tilde \ell } \left( \tilde \ell \!+\!1 \right) \left( \eta_S \!-\! \eta_{S'} \right)  \Theta \!\left( \eta_S \!-\! \eta_{S'} \right)  \right) \right] +  \left( S\leftrightarrow S' \right), \\
c_\ell^{(142)} \!\! &= & \!\! \frac{1}{8 \pi G a_S^2 \left( \bar \rho_S \!+\! \bar p_s \right)} \!\left( 1\! -\! \frac{\mathcal H_S^{-1}}{\Delta \eta_S}\right)\!\frac{1}{\Delta\eta_{S'}} \frac{2 \ell \left( \ell+1 \right) }{\pi \left( 2 \ell +1 \right)^2} \nonumber \\ 
&& \qquad \times \int \! dk \ k^4 P_\sigma \left( k \right) \dot T_k \left( \eta_S \right) \!\! \sum_{\tilde \ell = \ell\!-\!1, \ell \!+\!1 } \!\! j_{\tilde \ell  } \left( k \Delta \eta_S \right)  \int_{\eta_{S'}}^{\eta_O} \!\! d\eta\! \left( \eta_O \!- \!\eta \right)\! \left( \eta \!- \!\eta_{S'} \right)\! \dot T_k \left( \eta \right)j_{\tilde \ell} \left( k \Delta \eta \right)  \qquad +  \left( S\leftrightarrow S' \right) \nonumber \\
&\cong&  \!\! \frac{1}{8 \pi G a_S^2 \left( \bar \rho_S \!+\! \bar p_s \right)} \!\left( 1\! -\! \frac{\mathcal H_S^{-1}}{\Delta \eta_S}\right)\!\frac{1}{\Delta\eta_{S'}} \frac{2 \ell \left( \ell+1 \right) }{\pi \left( 2 \ell +1 \right)^2}  \!\! \sum_{\tilde \ell = \ell\!-\!1, \ell \!+\!1 } \!\!  \left[ \frac{I_{\tilde \ell}^2 \tilde \ell^2 }{\Delta\eta_S^3} P_\sigma \left( \frac{\tilde \ell }{\Delta \eta_S}\right) \dot T^2_{\tilde \ell /\Delta \eta_S}\! \left( \eta_S \right) \  \left( \!\eta_S \!-\! \eta_{S'}\! \right)\Theta \!\left( \!\eta_S \!-\! \eta_{S'}\! \right)   \right] \nonumber \\
&& +    \left( S\leftrightarrow S' \right)  .
\end{eqnarray}
\vspace{1cm}
\twocolumngrid

\section{Limber approximation}\label{app:lim}
In this work we have used the Limber approximation~\cite{lim} repeatedly. It approximates the integral of the 
product of a spherical Bessel function and a slowly varying function (e.g., a power law) by 
\begin{eqnarray} 
\int_{x_1}^{x_2} \!\!dx f\!\left( x \right) j_\ell\!  \left( x\right) \!\! &\cong&\!\!  I_\ell f\!\left( \ell \right)  \Theta\!\left( x_2 - \ell \right) 
  \Theta\! \left( \ell - x_1 \right)  \label{145c}
\end{eqnarray}
for $x_2> x_1$,where $\Theta$ denotes the Heaviside function defined by
\begin{eqnarray}
\Theta \left( x \right) = \left\{ \begin{array}{cc} 0, & x\le 0 \\ 1 , & x >0 \end{array} \right. ,
\end{eqnarray}
and $I_\ell^2 = 1.58/\ell$ describes the area under the first peak of the spherical Bessel function $j_\ell \left( x \right) $. 
This rather crude approximation considers the contribution under the first peak only, and it usually gives an overestimation, but never by more than a factor of $2$~\cite{BDG}. Of course, if the function $f$ varies heavily
in the region of the first peak, $\ell-1 <x<\ell+1$, the approximation cannot be used.

We are, in particular, interested in (note that $\Delta\eta=\eta_O-\eta$ and $\Delta\eta'=\eta_O-\eta'$)
\begin{eqnarray}\label{limb1}
\int_{{\eta_S}}^{{\eta_O}} \! \!  \! \!  &d\eta& \! \!   T \left(\eta \right) j_\ell \left( k \Delta\eta \right)  \cong \frac{1}{k} T \left( \eta_{\ell, k} \right) I_\ell \Theta \left( \eta_{\ell,k} - \eta_S \right), \\
\int_{{\eta_S}}^{{\eta_O}}\! \!   \! \!  &d\eta&\! \!  T \left( \eta \right) j'_\ell \left( k \Delta\eta \right) \label{limb2} \nonumber\\
&\cong& \frac{1}{k} T \left( \eta_{\ell-1, k} \right) I_{\ell-1} \Theta \left( \eta_{\ell-1,k} - \eta_S \right)  \nonumber \\
&-& \frac{1}{k} T \left( \eta_{\ell, k} \right) I_\ell \frac{\ell+1}{\ell} \Theta \left( \eta_{\ell,k} - \eta_S \right),  \\
\int_{{\eta_S}}^\eta\! \!  \! \!  &d\eta'& \frac{\dot T \left( \eta' \right)}{\left( k \Delta\eta' \right)^2} j_\ell \left( k \Delta\eta' \right) \label{limb3} \nonumber\\
&\cong&\frac{1}{k \ell^2} \dot T\! \left( \eta_{\ell, k} \right)\! I_\ell \Theta \left( \eta_{\ell,k} - \eta_S \right) \Theta \left( \eta - \eta_{\ell,k}   \right).  \\
\int_{{\eta_S}}^{{\eta_O}} \! \!  \! \! \!  \!   &d\eta& \! \!  \int_{{\eta_S}}^\eta d\eta' \frac{\dot T \left( \eta' \right)}{\left( k \Delta\eta' \right)^2} j_\ell \left( k \Delta\eta' \right)  \label{limb4} \nonumber\\
&\cong& \! \! \!  \!   \frac{1}{k^2\ell} \dot T \!\left(  \eta_{\ell, k} \right)\! I_\ell \Theta \left( \eta_{\ell,k} - \eta_S\right), \\
 \int_{{\eta_S}} ^{{\eta_O}} \!  \!  \! \!  \! \! &d\eta& \!  \!  \left( \eta-{\eta_S} \right)\left( {\eta_O} - \eta \right) \mathcal H \left( \eta \right) \dot T \left( \eta \right) \frac{j_\ell(k\Delta\eta) }{( k \Delta\eta )^2} \nonumber \\
& \cong &\!  \! \!  \! \frac{1}{\ell k^2} \!   \!  \left( \eta_{\ell,k} \!  - \!  \eta_S \right) I_\ell  \mathcal H  \!  \left( \eta_{\ell, k}\right) \dot T \!  \left( \eta_{\ell, k}\right) \Theta \left(  \eta_{\ell,k} \!  - \!  \eta_S\right),
 \qquad \label{limb5} \\
 \int_{{\eta_S}}^{{\eta_O}} \! \!  \! \!  &d\eta& \! \! \int_{{\eta_S}}^\eta \!  d\eta' T\left( \eta' \right) j'_{\ell} \left( k \Delta\eta' \right) \nonumber\\
 &\cong& \frac{\ell -1 }{k^2} T\left( \eta_{\ell -1 , k } \right) I_{\ell -1} \Theta\left(  \eta_{\ell-1,k} - \eta_S\right) \nonumber \\
 && \qquad - \frac{\ell +1 }{k^2} T\left( \eta_{\ell, k } \right) I_\ell \Theta  \left( \eta_{\ell,k} - \eta_S \right),  \\
 \int_{\eta_S}^{\eta_O}  \! \!  \! \!  &d\eta& \! \! \left( \eta \! - \! \eta_S \right) \left( \eta_O \! - \! \eta \right) T_k \! \left( \eta \right) j_\ell \left( k \Delta\eta \right) \nonumber\\
 & \cong &\!  \! \!  \! \frac{\ell}{ k^2} \!   \!  \left( \eta_{\ell,k} \!  - \!  \eta_S \right) I_\ell   T \!  \left( \eta_{\ell, k}\right) \Theta \left(  \eta_{\ell,k} \!  - \!  \eta_S\right),  \\
 \int_{\eta_S}^{\eta_O}  \! \!  \! \!  &d\eta& \! \! \left( \eta \! - \! \eta_S \right) \left( \eta_O \! - \! \eta \right) \dot T \! \left( \eta \right) j'_\ell \left( k \Delta\eta \right) \nonumber\\
  &\cong& \frac{\ell \! - \! 1}{k^2} I_{\ell \! - 1} \left( \eta_{\ell \! - \! 1,k} \! -\! \eta_S \right) \dot  T \! \left( \eta_{\ell \! - \! 1,k} \!\right)\Theta \! \left( \eta_{\ell \! - \! 1,k} \! -\! \eta_S \right)  \nonumber \\
&& \!\!\!\! - \frac{ \ell \! + \! 1}{k^2} I_{\ell}  \left( \eta_{\ell ,k} \! -\! \eta_S \right) \dot  T \! \left( \eta_{\ell ,k} \!\right)\Theta \! \left( \eta_{\ell ,k} \! -\! \eta_S \right).
\end{eqnarray}
We have used the following propriety of the spherical Bessel functions~\cite{CG}:
\begin{equation}
 j'_\ell \left( k \Delta\eta \right) = j_{\ell-1} \left( k\Delta\eta \right) - \frac{\ell+1}{k\Delta\eta} j_\ell\left( k\Delta\eta \right).
\end{equation}


\begin{thebibliography}{99}
\bibitem{benz} H. Nussbaumer and L. Bieri, {\em Discovering the Expanding Universe}, 
              Cambridge University Press (2009). 
\bibitem{SN1a} B. Schmidt et al., Astrophys. J. {\bf 507}, 46 (1998) [arXiv:astro-ph/9805200].\\
              A. Riess et al. Astron. J. {\bf 116}, 1009 (1998) [arXiv:astro-ph/9805201];\\ S. Perlmutter et al.,
              Astrophys. J. {\bf 517}, 565 (1999) [arXiv:astro-ph/9812133];\\              
              See also  {\small\tt http://www.nobelprize.org/nobel\_prizes/\\ 
              physics/laureates/2011/}.
\bibitem{SNnew}R. Amanullah et al.,  Astrophys. J. {\bf 716}, 712 (2010) [arXiv:1004.1711];\\ J. Guy et al. (2010),
              arXiv:1010.4743v1;\\ T. Holsclaw, et al., Phys. Rev. Lett. {\bf 105}, 241302  (2010) [arXiv:1011.3079];\\
              N. Suzuki et al. (2011)	arXiv:1105.3470v1.
\bibitem{Hui}L. Hui and P.B. Greene, Phys. Rev. {\bf D73}, 123526 (2006) [arXiv:astro-ph/0512159].
\bibitem{BDG}C. Bonvin, R. Durrer and M.A. Gasparini, Phys. Rev. {\bf D73}, 023523 (2006) [arXiv:astro-ph/0511183].
\bibitem{BDG.err}C. Bonvin, R. Durrer and M.A. Gasparini, Phys. Rev. {\bf D85}, 029901(2012) [arXiv:astro-ph/0511183].
\bibitem{BDK}C. Bonvin, R. Durrer and M. Kunz, Phys. Rev. Lett. {\bf 96}, 191302 (2006) [arXiv:astro-ph/0603240].
\bibitem{BenDayan} I. Ben-Dayan et al.,  JCAP {\bf 1204} (2012) 036 [arXiv:1202.1247].  
\bibitem{voids}K. Enqvist, Gen Rel. Grav. {\bf 40}, 451 (2008) [arXiv:0709.2044], and references therein.
\bibitem{walls}E. Di Dio, M. Vonlanthen, and R. Durrer,   JCAP {\bf 1202} (2012) 036 [arXiv:1111.5764].
\bibitem{Enea.master}E. Di Dio, Master thesis at ETHZ  (2010).
\bibitem{cris.vec}T. Hui-Ching Lu, K. Ananda and C. Clarkson, Phys. Rev. {\bf D77}, 043523 (2008) [arXiv:0709.1619].
\bibitem{cris.ten}K. N. Ananda, C. Clarkson and D. Wands, Phys. Rev. {\bf D75}, 123518 (2007) [arXiv:gr-qc/0612013v1].
\bibitem{2nd}E. Di Dio and R. Durrer, in preparation (2012).
\bibitem{Schmidt1}F. Schmidt and D. Jeong, arXiv:1204.3625.
\bibitem{Schmidt2}D. Jeong, F. Schmidt,  arXiv:1205.1512.
\bibitem{sachs}P. Schneider, J. Ehlers and E.E. Falco, {\em Gravitational Lenses}, (Springer Verlag, 1992).
\bibitem{EBD} G. F. R. Ellis, B. A. Bassett, P. K. Dunsby, Class. Quant. Grav. {\bf15} (1998) 2345 [arXiv:gr-qc/9801092].
\bibitem{ns}N. Straumann, {\em General Relativity with Applications to Astrophysics}, (Springer Verlag, Berlin, 2004).
\bibitem{RD}R. Durrer, Fund. Cos. Phys. {\bf 15}, 209 (1994) [arXiv:astro-ph/9311041]; \\
                    R. Durrer, {\em The Cosmic Microwave Background}, Cambridge University Press (2008).
\bibitem{DK} R. Durrer and T. Kahniashvili, Helv. Phys. Acta {\bf 71}, 445 (1997) [arXiv:astro-ph/9702226].
\bibitem{WMAP} D. Larsen et al., Astrophys.J.Suppl. {\bf 192}, 16 (2011) [arXiv:1001:4635].
\bibitem{Perlick} V. Perlick, Living Reviews in Relativity, {\bf 7},  9 (2004).
\bibitem{CG}M. Abramowitz and I. Stegun, {\em Handbook of Mathematical Functions}, 9th Printing, 
                    Dover Publications, New York (1970).
\bibitem{lim}M. LoVerde and N. Afshordi, Phys. Rev. {\bf D78}, 123506 (2008) [arXiv:0809.5112]. 
\end{thebibliography}
\end{document}